\documentclass[lettersize,journal]{IEEEtran}
\usepackage{amsmath,amsfonts}
\usepackage{algorithmic}
\usepackage{algorithm}
\usepackage{array}
\usepackage[caption=false,font=normalsize,labelfont=sf,textfont=sf]{subfig}
\usepackage{textcomp}
\usepackage{stfloats}
\usepackage{url}
\usepackage{verbatim}
\usepackage{graphicx}
\usepackage{cite}
\usepackage{booktabs}
\usepackage{multirow}

\usepackage{booktabs}
\usepackage{tabularx}
\usepackage{array}
\usepackage{siunitx}
\usepackage{xcolor}
\colorlet{blue}{black}
\usepackage{threeparttable}

\begin{document}

\title{Harmonia: Algorithm-Hardware Co-Design for Memory- and Compute-Efficient BFP-Based LLM Inference}

\author{Xinyu Wang,
        Jieyu Li,
        Yanan Sun,~\IEEEmembership{Senior Member,~IEEE,} 
        and Weifeng He,~\IEEEmembership{Senior Member,~IEEE}
\thanks{This paper was produced by the IEEE Publication Technology Group. They are in Piscataway, NJ.}
\thanks{Manuscript received February 23, 2026; revised July 30, 2026.}}

\markboth{Journal of \LaTeX\ Class Files,~Vol.~14, No.~8, August~2026}%
{Shell \MakeLowercase{\textit{et al.}}: A Sample Article Using IEEEtran.cls for IEEE Journals}


\maketitle

\hyphenpenalty=4000
\exhyphenpenalty=4000
\widowpenalty=0
\clubpenalty=0

\setlength{\floatsep}{0pt}
\setlength{\textfloatsep}{4pt}
\setlength{\abovecaptionskip}{0pt}
\setlength{\dbltextfloatsep}{4pt}

\begin{abstract}
Large Language Models (LLMs) are powerful but incur high memory and computation costs. Weight-only quantization is an effective solution with integer (INT) weights and floating-point (FP) activations being adopted to preserve accuracy. To further reduce the overhead of FP-INT arithmetic, prior works convert FP activations in linear layers into block floating point (BFP), where a group of activations shares a single exponent. However, prior works failed to extend BFP to attention layers due to significant accuracy loss, leading to suboptimal efficiency. To address this issue, we propose Harmonia, an algorithm-hardware co-design framework that enables all-layer BFP activations with a configurable hardware architecture. First, we systematically explore BFP configurations to achieve a better trade-off between accuracy and compression rate across all layers. Second, to further reduce KV-cache related storage and computation in attention layers, we propose an asymmetric bit-allocation strategy, combined with a hybrid offline-online outlier smoothing technique. Based on this, we aggressively compress the KV-cache from FP16 to 4b-mantissa BFP, with an average loss of less than 1\% on LongBench benchmark. Third, to fully exploit the benefits of all-layer BFP-activations, we propose a suite of dedicated hardware designs for Harmonia. Specifically, we propose a reconfigurable PE unit supporting mixed data formats (i.e., BFP-INT, BFP-BFP) and mixed data precisions, a real-time FP16-to-BFP converter for on-the-fly activation compression, and a flexible tiling-aware dataflow to reduce external memory traffic. We evaluate Harmonia in both linear layers and attention layers across eight widely used LLMs. Compared with prior works, experimental results show that Harmonia achieves 3.84$\times$ (up to 5.05$\times$) higher area efficiency, 2.03$\times$ (up to 3.90$\times$) better energy efficiency, and 3.08$\times$ (up to 4.62$\times$) speedup on average, respectively.
\end{abstract}


\section{Introduction}
\IEEEPARstart{L}{arge} Language Models (LLMs) \cite{touvron2023llama}, \cite{touvron2023llama2openfoundation}, \cite{jiang2023mistral7b}, \cite{zhang2022opt}, \cite{abdin2024phi}, \cite{achiam2023gpt}, \cite{liu2024deepseek} have emerged as a transformative foundation in artificial intelligence, demonstrating remarkable proficiency across a wide range of natural language processing tasks, including recommendation systems, chatbots, and personal agents. Guided by the scaling law \cite{kaplan2020scaling}, the most advanced LLMs’ parameters have grown to hundreds of billions, posing substantial demands on both memory and computational resources. For instance, deploying a Llama-3.1-405B model, which occupies around 800GB of memory, requires at least 10 high-end H100 GPUs. Additionally, during the generation phase, the autoregressive nature of LLMs necessitates recurrent retrieval of a large number of previously generated tokens, resulting in a memory bandwidth bottleneck \cite{lee2024infinigen}, \cite{kwon2023efficient}, \cite{sheng2023flexgen}, which severely impacts the performance and efficiency of model inference. These challenges emphasize an urgent need to improve the efficiency of LLM deployment.

\begin{figure}[t]
\centerline{\includegraphics[width=\linewidth]{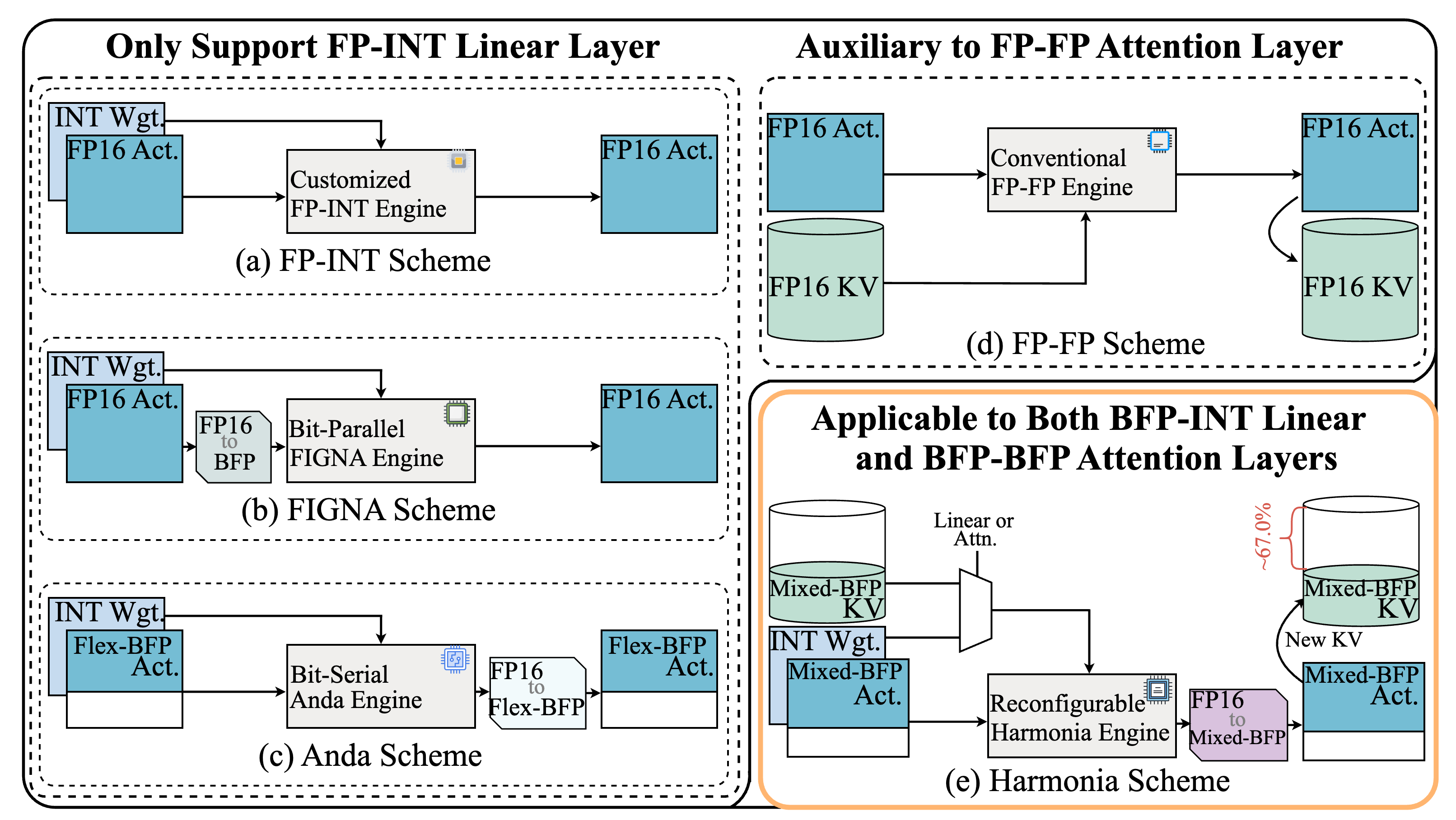}}
\caption{Comparison of different compute schemes: (a) customized FP-INT scheme, (b) FIGNA scheme, (c) Anda scheme, (d) conventional FP-FP scheme, and (e) the proposed Harmonia scheme.}
\label{fig1}
\end{figure}

Quantization is an effective approach for efficient model deployment. \cite{zhao2024atom}, \cite{xiao2023smoothquant}, \cite{ashkboos2024quarot}, \cite{lin2024duquant}, \cite{liu2024spinquant}, \cite{sun2024flatquant} quantize both floating-point (FP) weights and FP activations to low-precision integers. While weights can be statically quantized offline after training, the dynamic nature of activations necessitates runtime quantization, introducing additional latency and hardware costs \cite{hu2025m}, \cite{lee2024tender}, \cite{zhang2024mixpe}, \cite{kim2025oaken}. As a practical alternative, the weight-only quantization scheme \cite{frantar2022gptq}, \cite{dettmers2023case}, \cite{wang2023bitnet}, \cite{li2024norm}, \cite{lee2024owq}, \cite{dettmers2023spqr} where only weights are quantized to low bits (e.g., INT4), while activations remain in FP format (e.g., FP16), has been widely adopted. It can effectively reduce the model size while preserving the model's linguistic capabilities. To fully harness the computational gains from weight-quantized models, specialized execution units capable of processing FP-INT operands are typically required, as illustrated in Fig. \ref{fig1}(a). However, this scheme still involves costly FP-related operations (e.g., exponent alignment and normalization) and storage of FP-format activations, leading to significant hardware overhead.

To alleviate this issue, as shown in Fig. \ref{fig1}(b) and (c), FIGNA \cite{jang2024figna} and Anda \cite{fang2025anda} propose to convert FP activations involved in linear layers into the block floating point (BFP) format, where all elements within a block share a common exponent. This shared-exponent feature eliminates the need for costly per-element exponent alignment and normalization, enabling more efficient BFP-related multiplication and accumulation (MAC) on hardware.

\begin{figure}[t]
\centerline{\includegraphics[width=\linewidth]{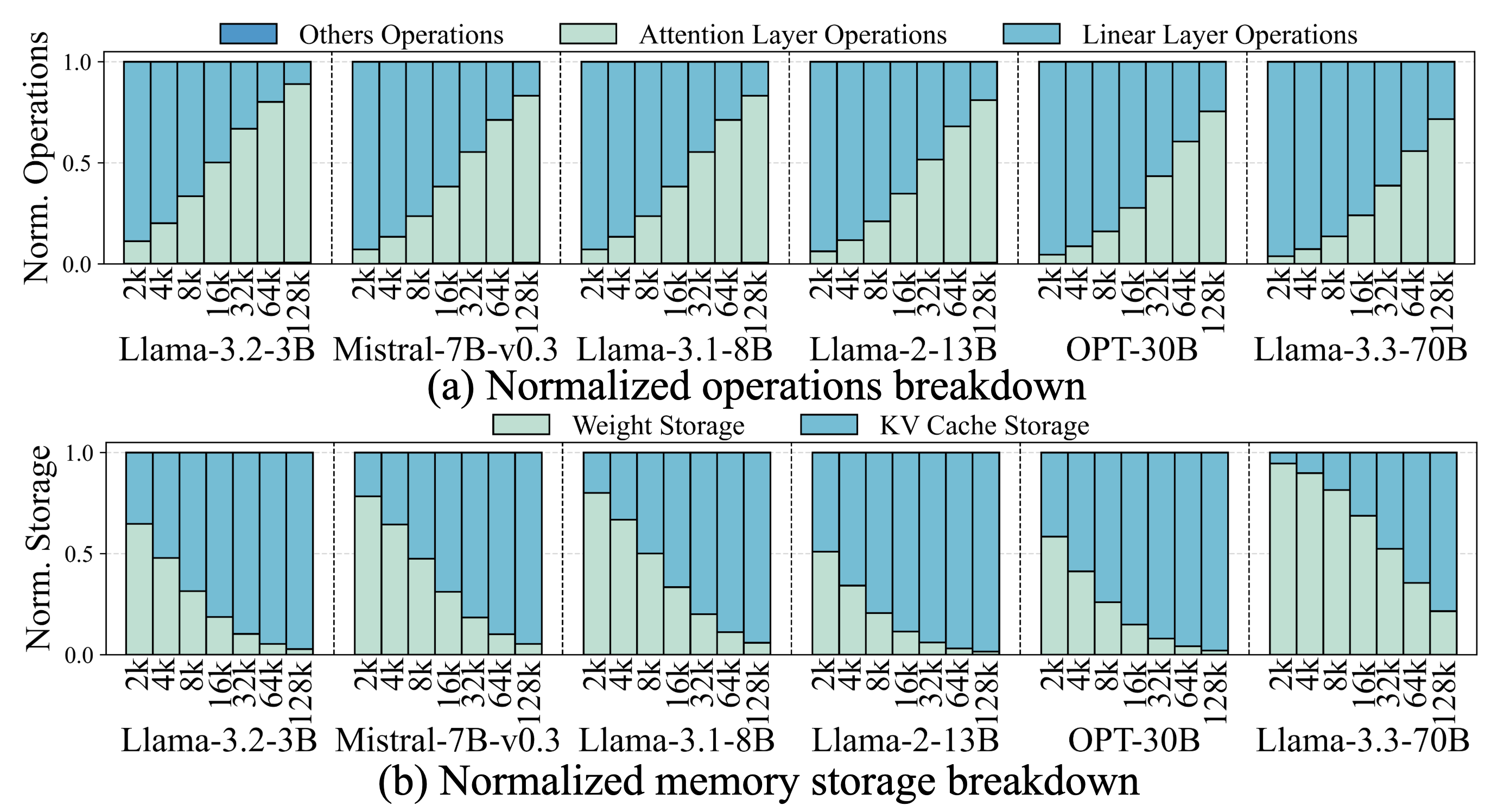}}
\caption{Normalized breakdown of (a) operations and (b) memory storage for models of different scales under varying sequence lengths.}
\label{fig2}
\end{figure}

However, these designs have two major limitations. First, they fail to extend the BFP format to FP activations in attention layers, which constitute a critical component of LLM architectures. As shown in Fig. \ref{fig2}(a), the dominant computation shifts between linear and attention layers depending on the input sequence length and model configuration. In short-context scenarios \cite{brown2020language}, \cite{li2025generation}, linear layers prevail across models of different scales, whereas in long-context applications \cite{bai2024longbench}, \cite{zhang2024chain} attention layers rapidly become the primary computational bottleneck as sequence length increases. Consequently, optimizing both layers is equally critical to achieving efficient and flexible adaptation across diverse workloads. Since both designs leave attention activations to be processed by conventional FP-based units shown in Fig. \ref{fig1}(d), the specialized arithmetic units optimized for linear layers remain underutilized during attention execution, leading to poor area efficiency and fragmented hardware resources.

Second, both designs retain the Key-Value (KV) cache in FP16 format, overlooking the frequent external memory accesses (EMA) it incurs. The KV cache stores intermediate attention keys and values, and can occupy over 90\% of the total memory footprint when the sequence is sufficiently long \cite{kwon2023efficient}. As shown in Fig. \ref{fig2}(b), its size scales linearly with sequence length, resulting in substantial EMA overhead in long-context scenarios. This excessive traffic becomes a major bottleneck during the decoding stage \cite{hooper2024kvquant}, \cite{lee2024infinigen}, \cite{hong2024flashdecoding++}, where computation is largely memory-bound and inference throughput is severely constrained. Therefore, maintaining the KV cache in full precision yields no tangible benefit in reducing EMA power or bandwidth pressure.

To overcome these limitations, as illustrated in Fig. \ref{fig1}(e), we propose Harmonia, an algorithm-hardware co-design framework that incorporates a unified BFP representation across activations in both linear and attention layers, supported by a configurable compute core capable of executing both BFP-INT and BFP-BFP operations. We begin by empirically exploring the feasibility of extending the BFP format to all-layer activations. Through experiments, we identify a BFP configuration that balances model accuracy and activation compression. Building upon this basis, we further propose an asymmetric bit-allocation strategy and a hybrid offline-online outlier smoothing technique to compress the KV cache from FP16 to 4-bit mantissas without noticeable accuracy loss. 

To fully leverage the benefits of the all-layer BFP-based activation representation, Harmonia integrates a suite of dedicated hardware units, including a reconfigurable PE unit for mixed data formats and precisions, a real-time FP16-to-BFP converter for on-the-fly activation compression, and a flexible tiling-aware dataflow to reduce the cost of EMA. Through these co-design approaches, Harmonia achieves a balanced trade-off among model accuracy, energy consumption, and computational efficiency.

In summary, our contributions are as follows:
\begin{itemize}
\item We employ a unified BFP format to represent activations across both linear and attention layers, adopting a well-selected configuration with a group size of 32 and 8-bit mantissas for all activations except the KV cache.
\item We propose an asymmetric bit allocation strategy and an offline-online hybrid outlier smoothing technique that together compress the KV cache to about 33.0\% of its original size with negligible accuracy loss.
\item We design a reconfigurable PE unit that unifies BFP-INT and BFP-BFP operations, achieving up to 4.85$\times$ higher area efficiency and 4.52$\times$ better energy efficiency over baselines.
\item Based on the above techniques, we present Harmonia, an LLM accelerator that dynamically compresses activations and supports flexible arithmetic across heterogeneous precisions and formats. Compared with SOTA accelerators, Harmonia achieves up to 5.05$\times$ higher area efficiency, 3.90$\times$ higher energy efficiency, and 4.62$\times$ speedup, with average gains of 3.84$\times$, 2.03$\times$, and 3.08$\times$, respectively.
\end{itemize}

\begin{figure}[tbp]
\centerline{\includegraphics[width=\linewidth]{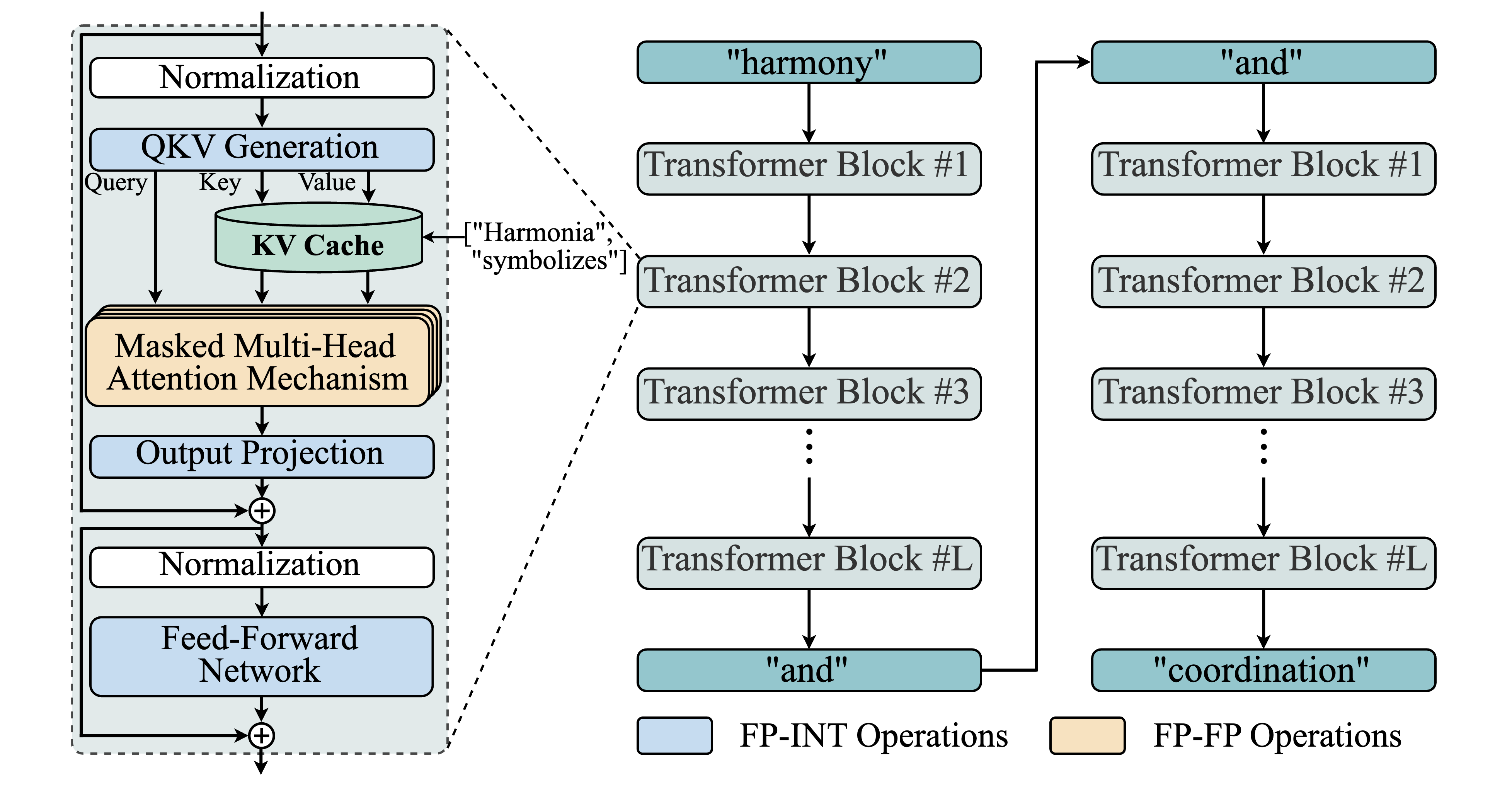}}
\caption{Illustration of the weight-only quantized LLM architecture.}
\label{fig3}
\end{figure}

\section{Background and Motivation}
This section first outlines the inference paradigm of weight-only quantized LLMs, followed by an introduction to the BFP format and its existing applications. After that, we further analyze the limitations of prior works, which reveal opportunities and motivate our design.

\subsection{Weight-only Quantized LLMs}
As shown in Fig. \ref{fig3}, LLMs are built upon the Transformer architecture \cite{vaswani2017attention}, which stacks multiple Transformer blocks composed of several linear layers and a masked self-attention mechanism. The generative inference of LLMs consists of two stages: prefilling and decoding. In the prefill stage, the prompt tokens are first processed through three linear layers to produce the Query (Q), Key (K) and Value (V) matrices. The attention scores are then derived from Q and K, which are subsequently used to perform a weighted sum over the V matrix. The intermediate results generated by the attention block are then passed through a feed-forward network to produce the final output. Meanwhile, the K and V matrices produced in this stage are stored as the KV cache, which serves to avoid redundant computation in the subsequent decode stage. During decoding, the newly generated token, together with the cached K and V, is used to generate the next token and update the KV cache repeatedly until a termination condition is met.

To reduce the costly memory footprint of LLM deployment, quantization has been widely applied to weights in linear layers. In this scheme, weights are quantized to low bit widths while activations remain in FP format. Compared to quantizing both weights and activations, weight-only quantization can compress model parameters to ultra-low bit widths with negligible accuracy loss, facilitating efficient deployment of LLMs on resource-constrained devices in edge computing scenarios.

Despite its advantages, weight-only quantization exhibits notable limitations in both computation and memory aspects. In terms of computation, quantized linear layers introduce mixed-format FP-INT operations, while attention layers continue to rely on FP-FP operations. Both operation types involve expensive exponent alignment and normalization \cite{jang2024figna}, which complicate hardware implementations and limit efficiency gains. In terms of memory, the KV cache is still maintained in FP format. As illustrated in Fig. \ref{fig2}(b), when the sequence length increases, the KV cache gradually dominates memory storage, overtaking model weights as the primary memory bottleneck during inference.

Consequently, optimizing FP activations across all layers emerges as a key opportunity for improving the overall efficiency of weight-only quantized LLMs.

\subsection{Block Floating Point}
\begin{figure}[t]
\centerline{\includegraphics[width=\linewidth]{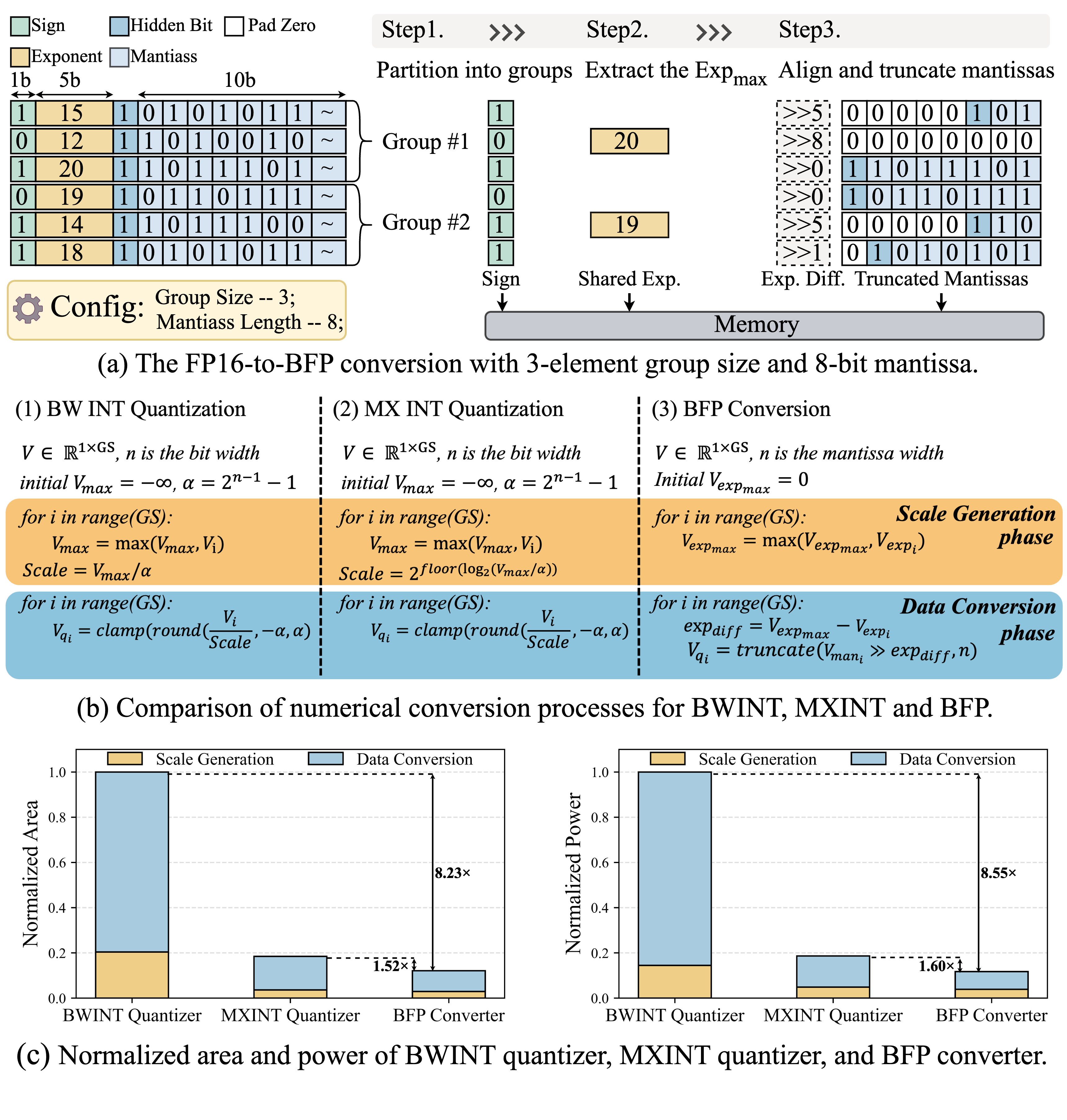}}
\caption{Illustration of (a) the FP16-to-BFP numerical conversion procedure, (b) a comparison of the BWINT, MXINT, and BFP conversion processes, and (c) the normalized hardware area and power breakdown of the corresponding implementations.}
\label{fig4}
\end{figure}

To reduce the computational and storage overhead of FP activations without significantly degrading model performance, Block Floating Point (BFP) \cite{drumond2018training} has been proposed as a balanced trade-off between the wide dynamic range of traditional FP formats and the simplicity of INT formats. Fig. \ref{fig4}(a) shows the steps to convert an FP16 vector to BFP: (1) partition the vector into groups, (2) extract the largest exponent in each group as the shared exponent, and (3) right-shift and truncate mantissas based on exponent differences. For comparison, Fig. \ref{fig4}(b) and Fig. \ref{fig4}(c) illustrate the numerical conversion procedures and corresponding area and power costs of block-wise (BW) INT quantization and microscaling (MX)\cite{rouhani2023ocp} INT quantization, respectively. Owing to its simplified conversion process, BFP significantly reduces the overhead of online numerical conversion. Compared to BWINT and MXINT quantizers, the BFP converter reduces the area consumption by 8.23$\times$ and 1.52$\times$, and achieves power savings of 8.55$\times$ and 1.60$\times$, respectively.

The BFP format further provides flexibility by allowing different group size and mantissa bit widths, enabling a tunable trade-off between model accuracy and hardware efficiency. In general, smaller groups with sufficient mantissa bits yield lower error, while larger groups and more aggressive mantissa truncation increase the error, which can degrade model accuracy. Consequently, selecting an appropriate BFP configuration to balance model accuracy and efficiency remains a non-trivial challenge.

To mitigate the impact of numerical errors on model accuracy, one feasible strategy is BFP-aware training \cite{zhang2022fast},\cite{darvish2023shared}, \cite{guo2023boost}, which directly adopts the BFP format during training. However, LLM training is prohibitively expensive \cite{chowdhery2023palm}, \cite{hoffmann2022training}, \cite{jiang2024megascale}, requiring massive computation and time, which limits its practicality. Instead, another line of work introduces BFP conversion at inference time. FIGNA \cite{jang2024figna} adopts a conservative strategy by analytically relating mantissa bit width to numerical precision and extending the mantissa to match FP-based results. Although this preserves accuracy, storing activations in FP16 format eliminates both memory and bandwidth savings. Anda \cite{fang2025anda} employs a customized search algorithm to allocate variable mantissa bit widths to different types of activations in linear layers under bounded accuracy loss, aiming to minimize total bit operations.

However, prior efforts mainly target linear layers, neglecting FP-FP operations between activations in attention layers and the substantial storage overhead of the KV cache. To fully exploit the potential of the BFP format, it is essential to explore the feasibility of converting all activations involved in LLM inference into the BFP format and, on this basis, aggressively compress the KV cache to minimize storage cost.

\subsection{Opportunities and Challenges}
\label{sec2-C}
\begin{figure}[tbp]
\centerline{\includegraphics[width=\linewidth]{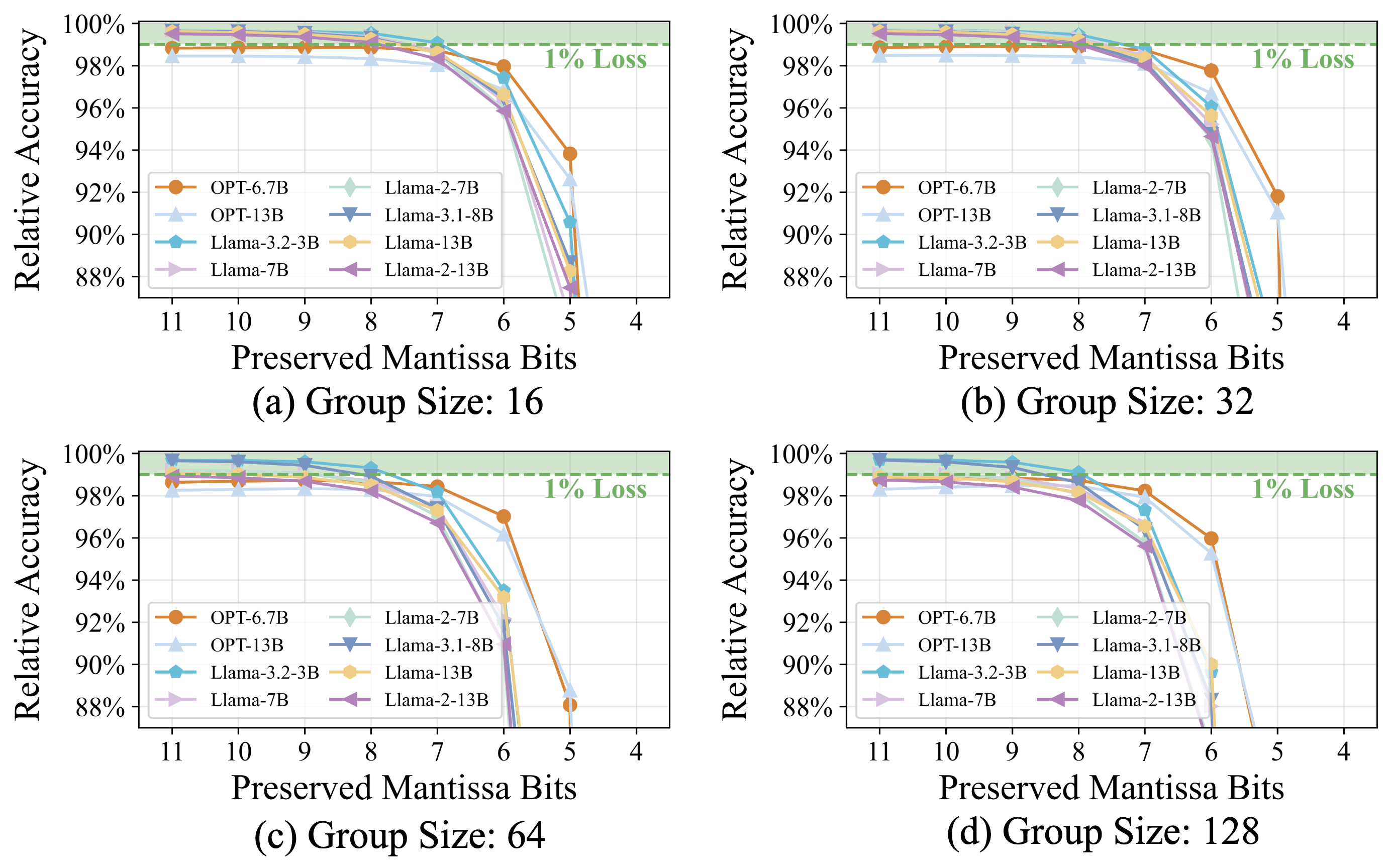}}
\caption{Relative accuracy of different models under various preserved mantissa bits and group sizes.}
\label{fig5}
\end{figure}

In this work, we investigate the possibility of converting all FP16 activations into the BFP format, including those involved in linear layers as well as the Q, K, V representations and attention scores within the attention mechanism. To assess the impact of different BFP configurations on model accuracy, we adopt perplexity on the WikiText2 dataset \cite{merity2016pointer} as the evaluation metric and evaluate eight widely used pre-trained models as benchmarks. The perplexity of the full-precision model serves as the 100\% baseline, and the relative performance across various mantissa bit widths and group sizes is shown in Fig. \ref{fig5}.

The results indicate that model accuracy is highly sensitive to the mantissa bit width. Each subplot of Fig. \ref{fig5} demonstrates that different models exhibit varying levels of accuracy degradation under the same BFP configuration, but they share a consistent trend: accuracy decreases as the mantissa bit width shrinks. Notably, when the bit width drops below 8 bits, accuracy declines sharply. Using 8-bit mantissas keeps the accuracy degradation around 1\% for most models, which we consider an acceptable loss. Therefore, all subsequent analyses and experiments adopt 8-bit mantissas for activations except those in the KV cache.

Furthermore, by comparing different subplots in Fig. \ref{fig5}, we observe that the group size also plays a crucial role in determining model accuracy. Larger groups achieve a higher activation compression rate but amplify accuracy loss caused by truncated mantissas. To balance accuracy and memory efficiency, we adopt a group size of 32 for all activations under the 8-bit mantissa configuration.

\begin{figure}[tbp]
\centerline{\includegraphics[width=\linewidth]{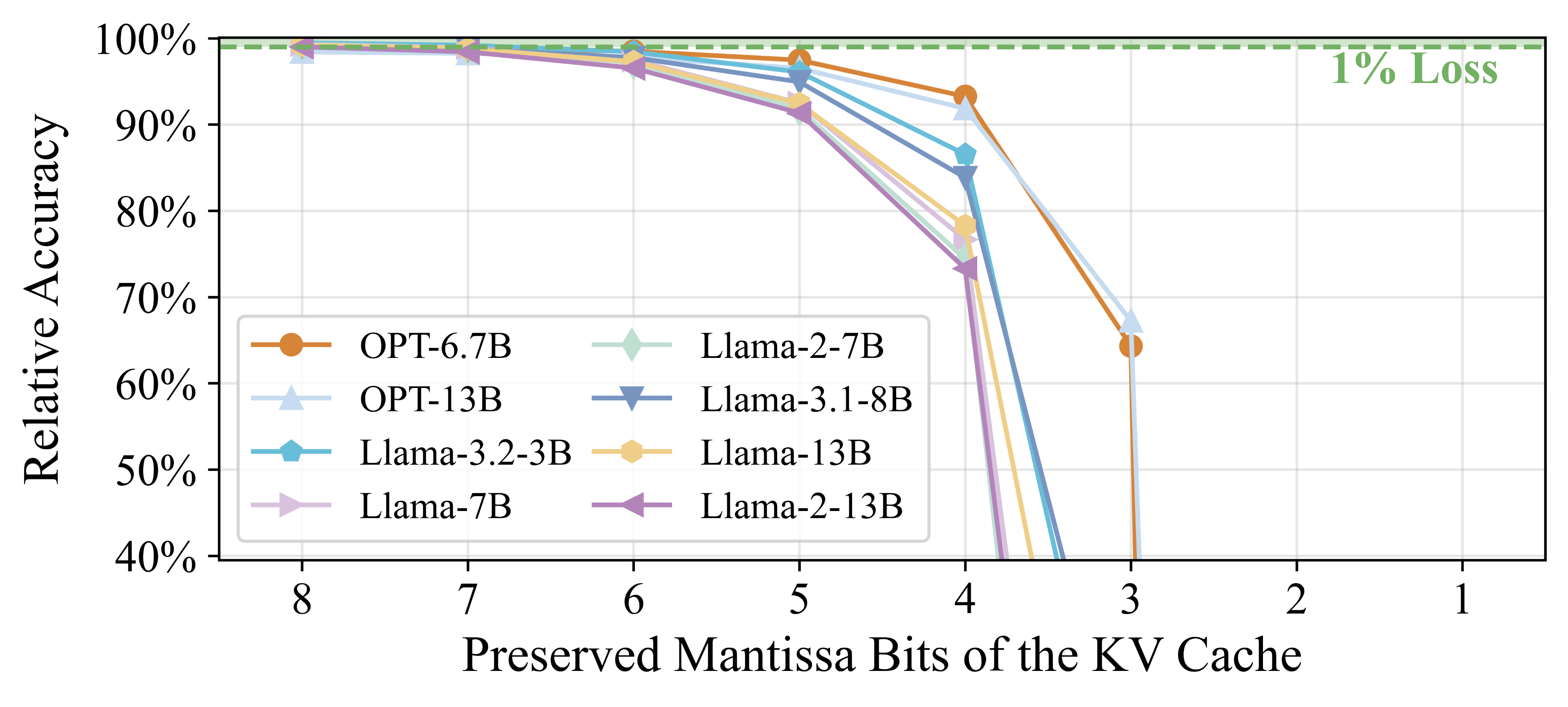}}
\caption{Relative accuracy of different models under various preserved mantissa bits of the KV cache.}
\label{fig6}
\end{figure}

To further mitigate the memory overhead of the KV cache, we evaluate the impact of truncating its mantissas to lower bit widths. In this experiment, the group size is fixed at 32, and the mantissa length of other activations is set to 8. We then gradually reduce the KV cache mantissa precision from 8 to 1 bit to evaluate accuracy degradation with decreasing bit width. As shown in Fig. \ref{fig6}, accuracy deteriorates progressively with aggressive truncation and drops sharply below 5 bits. For most models, the relative accuracy falls below 90\%, and in some cases approaches 70\% at 4 bits. These results highlight the challenge of compressing the KV cache to ultra-low bit widths while maintaining acceptable accuracy, motivating the need for advanced strategies to balance precision and efficiency.

In summary, our study reveals two key insights. \noindent\textbf{First (Opportunity)}, with a well-chosen combination of group size and mantissa length, model accuracy can be maintained within an acceptable margin. \noindent\textbf{Second (Challenge)}, compressing the KV cache to ultra-low bits remains difficult and causes severe accuracy loss, hindering practical deployment. These findings motivate us to extend the BFP format to all activations in LLMs to fully exploit its potential. In addition, it highlights the need for advanced optimization strategies to mitigate the accuracy loss incurred by aggressive KV-cache compression.

\section{Harmonia Algorithm}
\label{sec3}
This section presents the algorithmic foundation of Harmonia. Based on the experimental results in Sec. \ref{sec2-C}, we set the group size to 32, adopt a 5-bit shared exponent per group, and use an 8-bit mantissa for activations except the KV cache. In this section, we begin by introducing a dynamic BFP conversion framework, which converts activations to the BFP format during inference. To mitigate the accuracy loss caused by compressing the KV cache to 4-bit mantissas, we propose two key optimizations: (1) an initial-local asymmetric bit allocation strategy and (2) an offline–online hybrid outlier smoothing technique.

\subsection{BFP Conversion Framework}
\begin{figure}[tbp]
\centerline{\includegraphics[width=\linewidth]{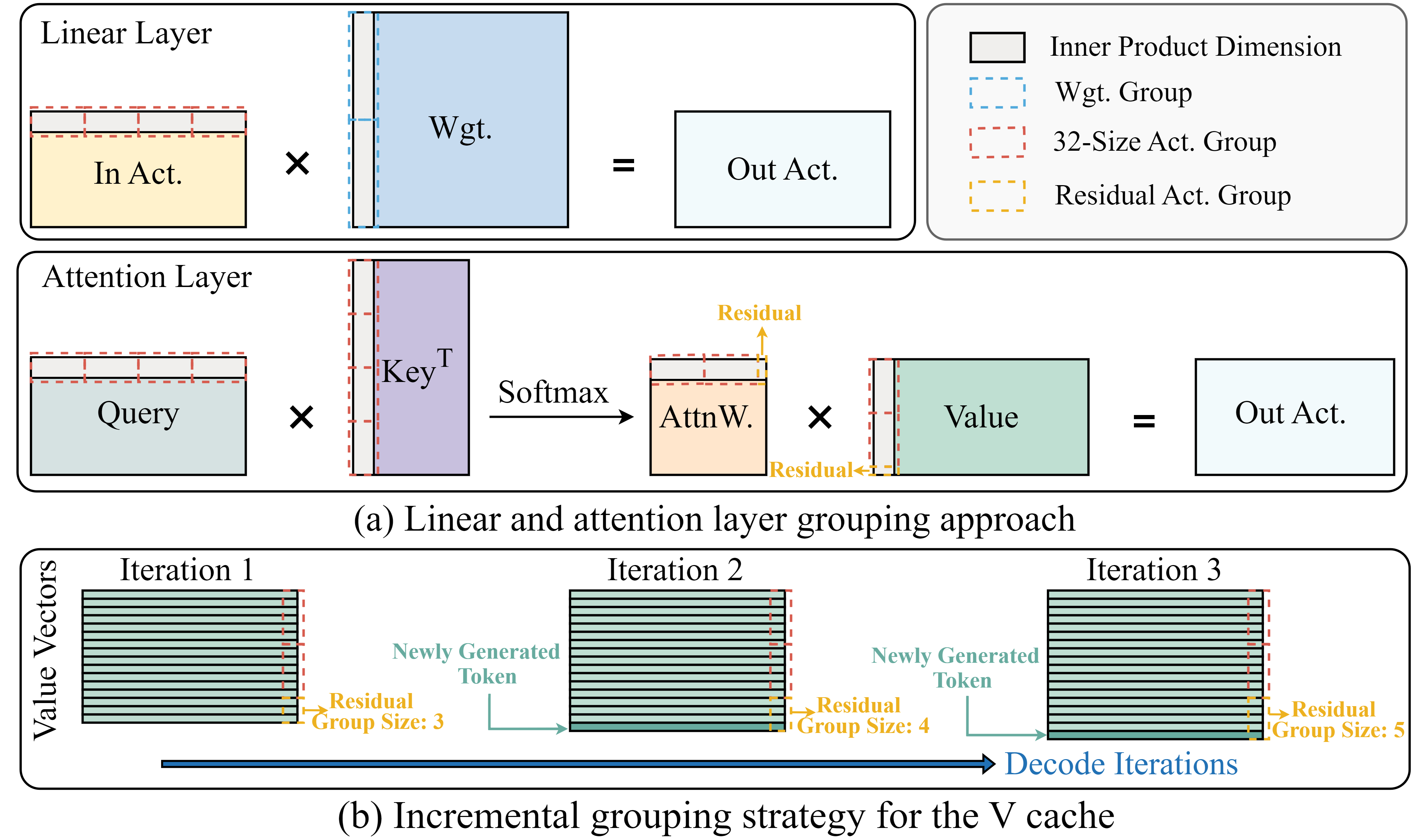}}
\caption{(a) Grouping strategies for different activation types. (b) Incremental grouping strategy applied to the V cache.}
\label{fig7}
\end{figure}

Given the dynamic nature of activations, online conversion to the BFP format at runtime is essential. To fully exploit BFP’s shared-exponent property, activation grouping must align with the inner-product computation direction. As shown in Fig. \ref{fig7}(a), activations in both linear and attention layers are grouped per token, except for the Value (V) matrix. This grouping strategy aligns well with the streaming pattern of the inference process, because the hidden dimension is an integer multiple of the group size in both the prefill and decode stages.

In contrast, for V, the dynamic growth of the sequence length disrupts group completeness, meaning that the most recent tokens may not form a full group. We refer to this partial block as a residual group. To address this issue, we propose an incremental grouping strategy for BFP conversion of V, as illustrated in Fig. \ref{fig7}(b). Instead of waiting for a full group to form, each newly generated token is iteratively appended to the residual group, which is temporarily converted based on its current size to produce intermediate results. The residual group is updated at every iteration as a new token arrives. Once the group reaches the predefined size, a final BFP conversion is performed, and the result is committed to the V cache.

This strategy ensures that all MACs are performed using BFP-formatted activations, thereby avoiding the requirement to introduce additional hardware units dedicated to handling the V vectors in the residual group and streamlining the overall datapath for a more compact and efficient design.

\subsection{Initial-Local Asymmetric Bit Allocation}
\label{sec3-B}
\begin{figure}[tbp]
\centerline{\includegraphics[width=\linewidth]{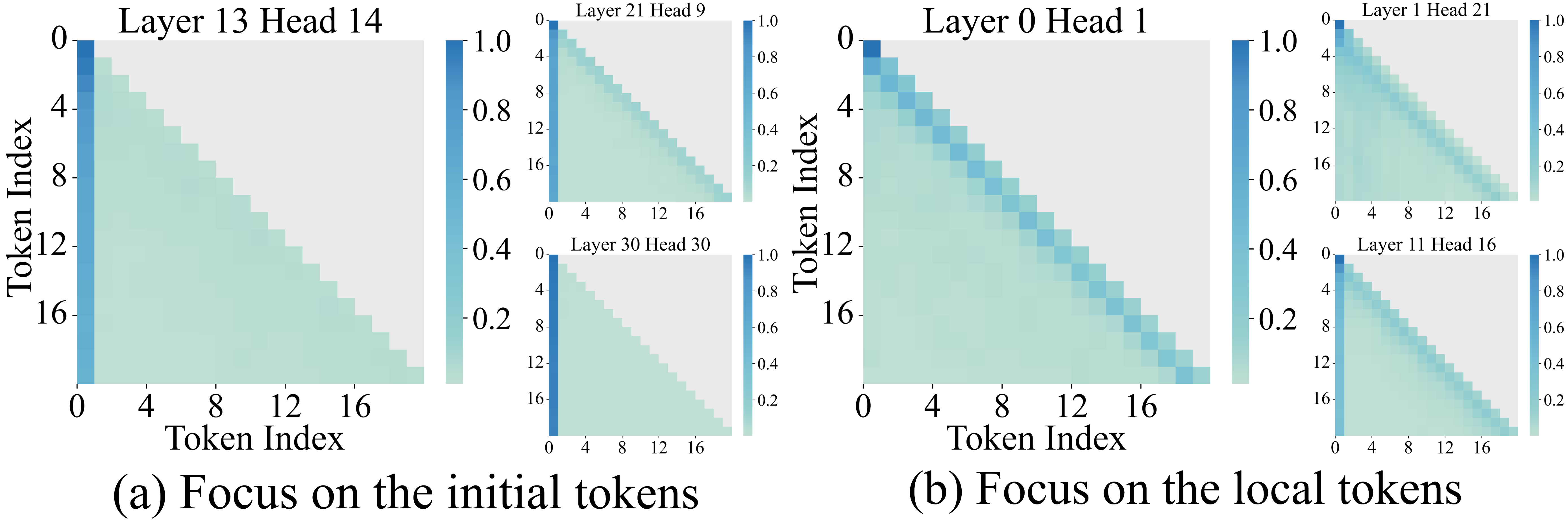}}
\caption{Average attention scores in Llama-3.1-8B over 128 sentences, each with a length of 20: (a) initial-token attentive heads, (b) local-token attentive heads.}
\label{fig8}
\end{figure}

To address the challenge outlined in Sec. \ref{sec2-C} of recovering model accuracy when truncating the KV cache mantissas to extremely low bit widths, we begin by analyzing the attention score distributions across different models. As illustrated in Fig. \ref{fig8}, this analysis reveals an important phenomenon: the model exhibits a strong tendency to focus on the initial few tokens and the most recent tokens, indicating a clear temporal pattern in the attention distribution. This observation reveals that model accuracy is particularly sensitive to the precision of these initial and local tokens, as they play a pivotal role in guiding the generation process.

Motivated by this insight, we propose an asymmetric bit allocation strategy that adaptively assigns higher precision to these critical regions of the sequence. Specifically, the mantissas of the initial and local tokens in the KV cache are allocated 8-bit precision, while those of others are truncated to 4 bits. The numbers of the initial and local tokens are 32 and 64, respectively. Under a 4K sequence length, 97.6\% of the KV cache adopts a 4-bit mantissa while only 2.4\% uses an 8-bit mantissa, reducing the KV-cache storage by 3.05×.

\begin{figure}[tbp]
\centerline{\includegraphics[width=\linewidth]{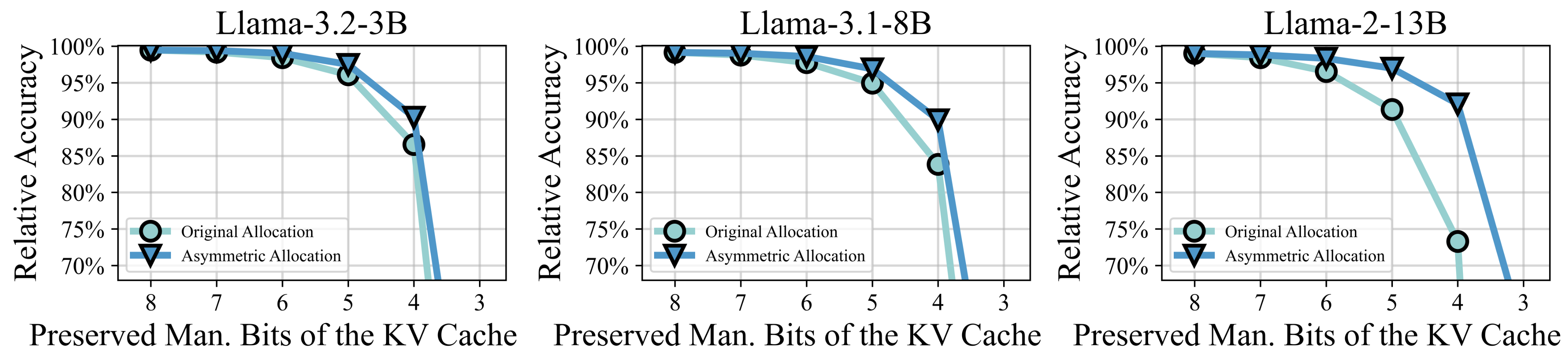}}
\caption{Accuracy improvement across three models achieved by applying the asymmetric bit-allocation strategy.}
\label{fig9}
\end{figure}

To validate the effectiveness of this strategy, Fig. \ref{fig9} compares three representative models with and without the asymmetric bit allocation. As shown in the figure, our approach achieves notable accuracy improvements, with a relative accuracy gain of 9.54\% across three models on average, effectively mitigating the precision loss caused by low-bit mantissas of the KV cache. The results show that this strategy preserves the attention fidelity of contextually important tokens while substantially reducing computation and memory overhead, achieving a favorable balance between accuracy and efficiency.

\subsection{Offline-Online Hybrid Outlier Smoothing}
\begin{figure}[tbp]
\centerline{\includegraphics[width=\linewidth]{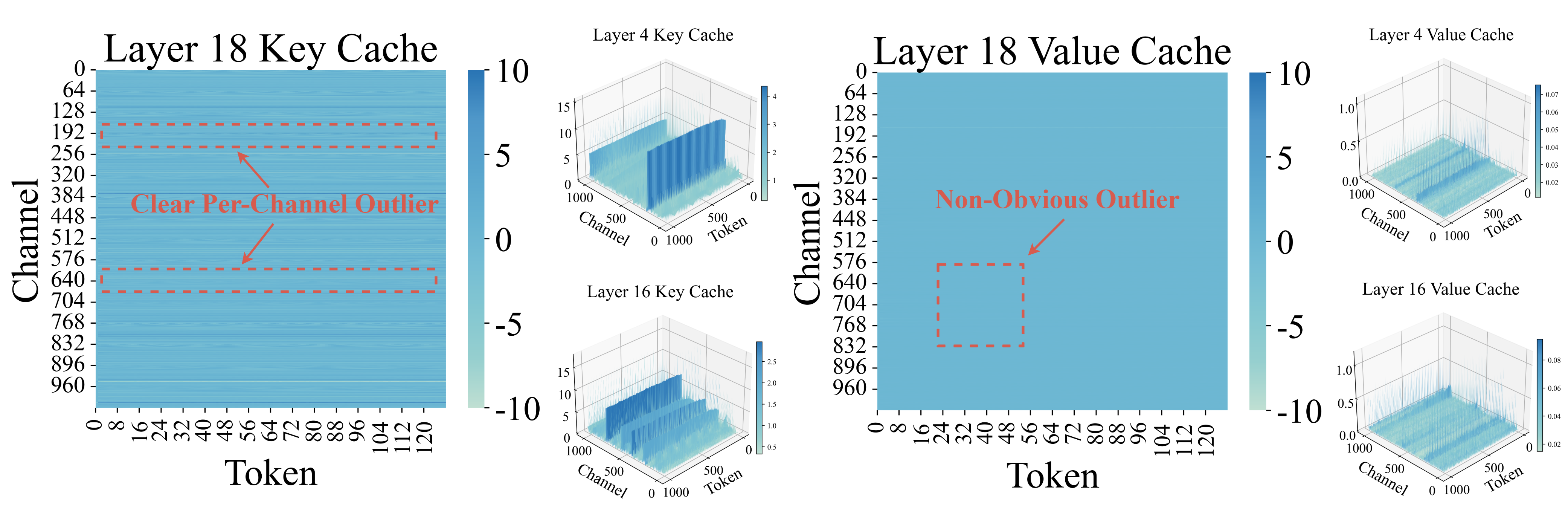}}
\caption{Magnitude distribution of values in the KV cache of Llama-3.1-8B.}
\label{fig10}
\end{figure}

Although our asymmetric bit-allocation strategy can improve the accuracy, it still poses notable accuracy loss compared to the BFP with the 8b-mantissa scheme. We have found that this is because of the presence of outliers. We analyze the magnitude distribution of KV cache values in Fig. \ref{fig10}. The V matrix exhibits a relatively uniform distribution, while K shows clear channel-wise outliers. When K is grouped across channels (i.e., along the token dimension) for BFP conversion under our grouping strategy, these outliers dominate the shared exponent, forcing smaller values to zero. The resulting conversion errors propagate through the attention scores and further to the outputs, ultimately degrading accuracy. To mitigate this, we propose an offline-online hybrid outlier smoothing technique that suppresses such outliers. After outlier smoothing, the magnitude distribution will be more concentrated, which makes it easier to realize the 4-b mantissa truncation without large accuracy loss.

For offline smoothing, we introduce a per-channel scaling mechanism to smooth outliers in K. We define a scaling factor $S \in \mathbb{R}^{1 \times C}$, where $C$ denotes the hidden dimension of K. Each channel of K is multiplied by its corresponding scaling factor $s_c$, while the corresponding channel of Q is divided by $s_c$ to preserve the original results, which can be described as:
\begin{equation}
\label{eq1}
P = softmax(QK^T) = softmax((Q \oslash S) \cdot (K^T \odot S))
\end{equation}

To implement this offline smoothing, there are two key challenges. First, dynamically applying per-channel scaling to Q and K during inference complicates the datapath and increases the computation overhead at runtime. Second, determining appropriate scaling factors remains non-trivial. To address the first issue, we observe that the Q and K are derived from the input activations through linear projections. This provides an opportunity to absorb the scaling factors into the weights of linear layers, thereby eliminating the need for explicit scaling during inference. Under this transformation, Eq. (\ref{eq1}) can be reformulated as:
\begin{equation}
\label{eq2}
P = softmax((X \cdot (W_Q \oslash S)) \cdot (({W_K}^T \odot S) \cdot X^T)) 
\end{equation}

To address the second issue, different from activation-aware weight quantization approaches, such as SmoothQuant \cite{xiao2023smoothquant} and AWQ \cite{lin2024awq}, which hand-craft the scaling factors, we treat the scaling factors as learnable parameters. This allows the scaling to be adaptively optimized, enabling a better match to the characteristics of different models. Specifically, we employ a calibration dataset to determine the optimal $S$ that minimizes the mean square error (MSE) between the original block outputs and the outputs produced when applying the scaling to Q and K. The optimization objective can be formulated as: 
\begin{equation}
\label{eq3}
S = \arg\min_{S} \bigl\| \mathcal{F}(W, X) - \mathcal{F}\bigl(W, \mathrm{Convert}_{\mathrm{BFP}}(X); S\bigr) \bigr\|_{2}^{2}
\end{equation}

\begin{figure}[tbp]
\centerline{\includegraphics[width=\linewidth]{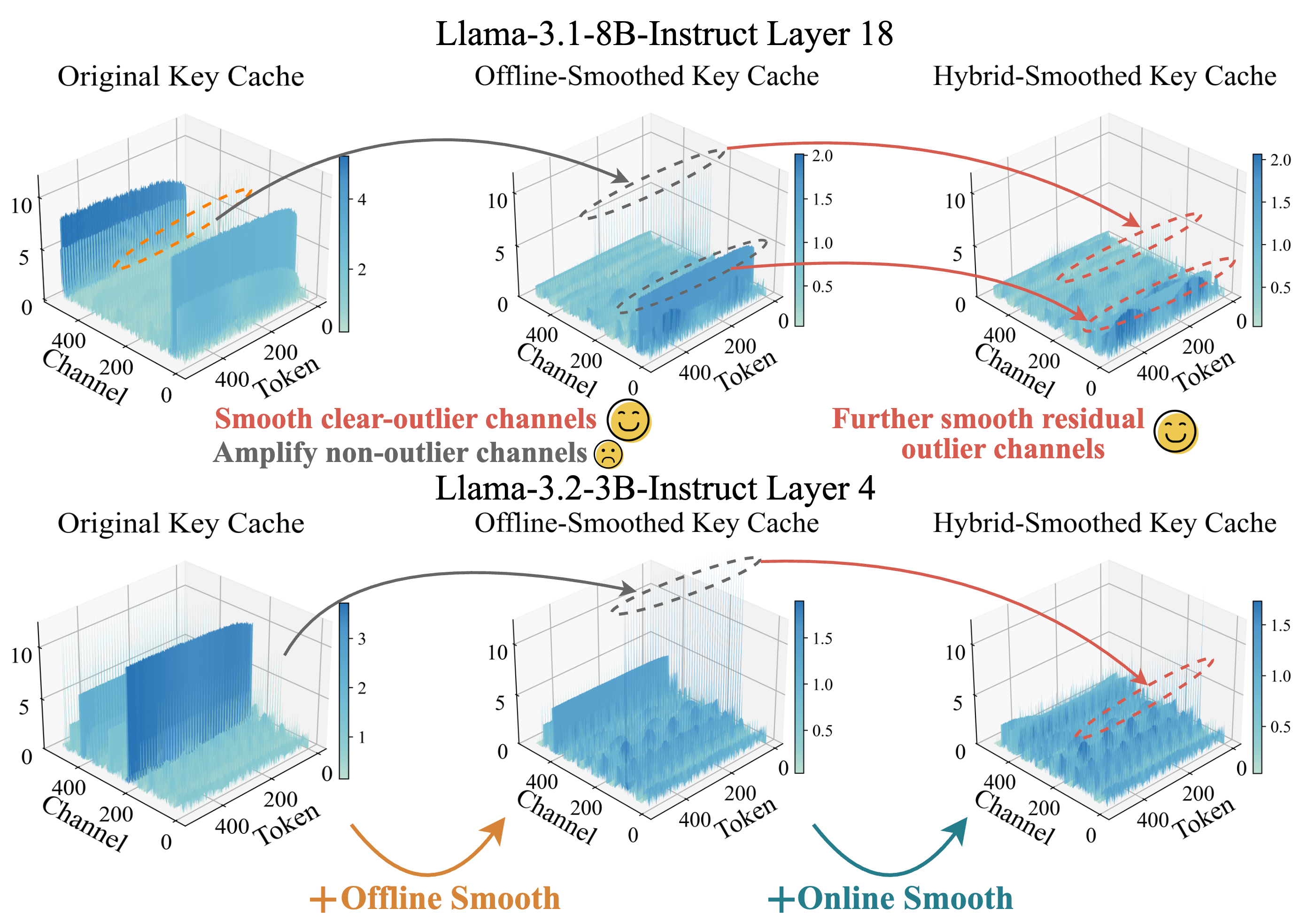}}
\caption{Effect of outlier suppression using the offline-online hybrid smoothing strategy.}
\label{fig11}
\end{figure}

where $\mathcal{F}$ represents a transformer block function, $W$ and $X$ are full precision weights and activations, $\mathrm{Convert}_{\mathrm{BFP}}$  represents the BFP converter, and $S$ is a learnable vector of scaling factors. As illustrated in Fig. \ref{fig11}, after applying our scaling-based smoothing strategy, the channel-wise outliers are effectively suppressed, and the overall distribution of K becomes more uniform, indicating improved numerical stability for subsequent BFP conversion. 

Notably, the results in Fig. \ref{fig11} also highlight an additional opportunity for improvement: while outliers are suppressed, the values in previously non-outlier channels are amplified, leading to a wider overall value range and introducing potential numerical instability. To mitigate this, we apply the online smoothing, i.e., an incremental optimization that adaptively handles dynamically changing distributions during inference. As shown in Fig. \ref{fig10}, K exhibits a clear intra-channel similarity across tokens. Additionally, because of the shift-invariance property of the softmax operation, subtracting a per-channel offset from K does not affect the resulting attention scores. Leveraging these two properties, we perform online smoothing on a subset of K channels with prominent outliers to stabilize their value ranges. To minimize runtime overhead, we adopt a lightweight offset selection strategy. Specifically, we first identify the maximum absolute value for each channel within the initial 32-token window. Then, we select the top-k channels with the largest magnitudes and assign half of each as the corresponding channel offset, while setting the offset of the remaining channels to zero. As illustrated in Fig. \ref{fig11}, applying online smoothing effectively suppresses outlier-induced divergence.

\begin{figure}[tbp]
\centerline{\includegraphics[width=\linewidth]{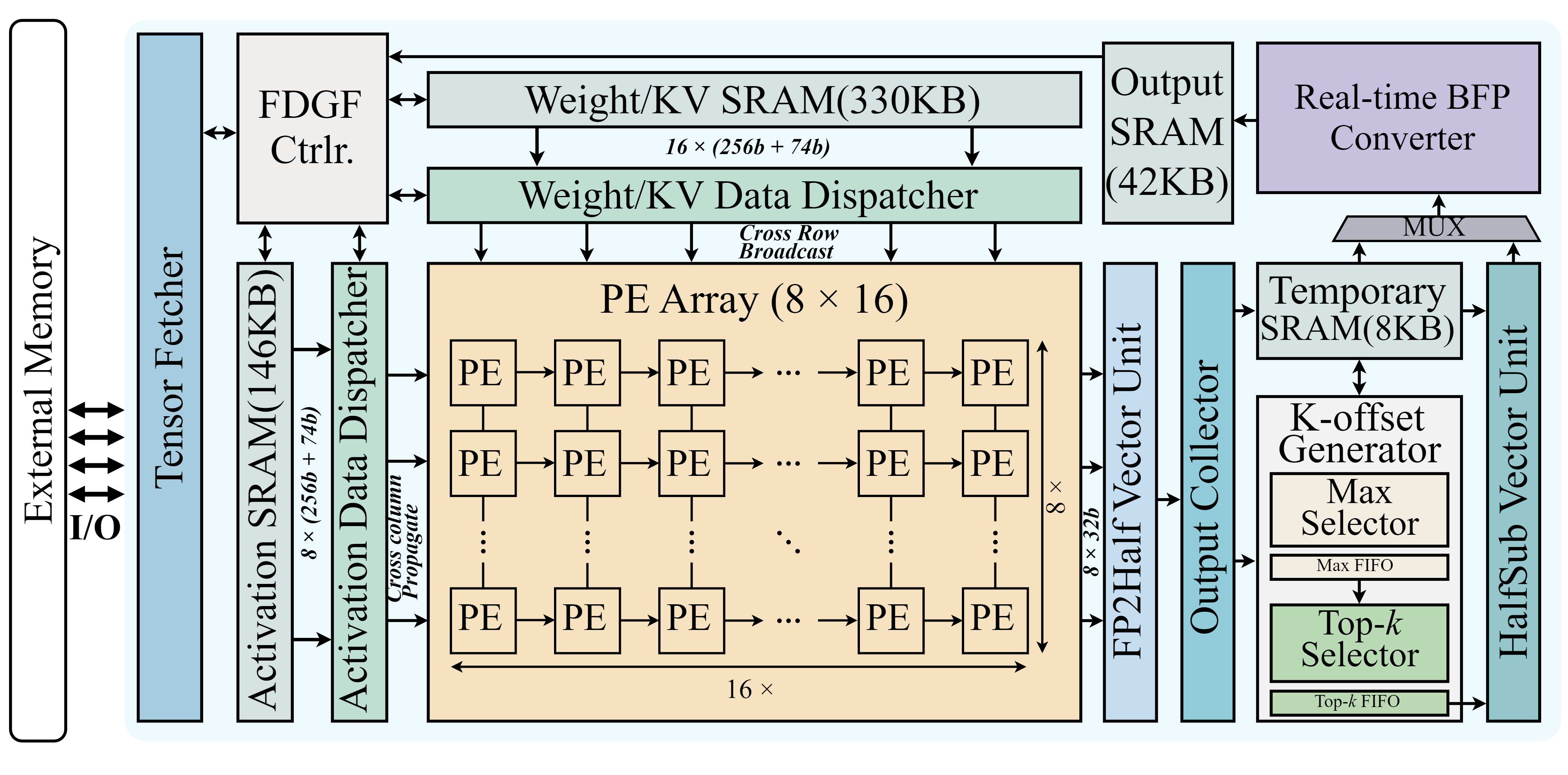}}
\caption{The overall architecture of Harmonia.}
\label{fig12}
\end{figure}

In summary, the proposed offline-online hybrid smoothing effectively suppresses outliers, achieving a balance between numerical stability and low runtime hardware overhead. This approach lays a solid foundation for simple and efficient hardware implementations in large-scale LLM inference.

\section{Harmonia Architecture}
\label{sec4}
This section presents the Harmonia architecture, designed to accelerate operations in both linear and attention layers. We first outline the overall Harmonia system, then detail its three key components: (1) a reconfigurable PE that supports MACs across heterogeneous data precisions and formats, (2) a real-time BFP converter for on-the-fly activation conversion, and (3) a tiling-aware dataflow that optimizes memory access.
\subsection{Overview}
Fig. \ref{fig12} illustrates the overall Harmonia architecture, which integrates a reconfigurable PE array, a real-time BFP converter, a dataflow controller, an online channel-wise K-offset generator, high-density dual-port on-chip SRAMs, data dispatchers, and vector units. During runtime, the computation proceeds as follows:

(1) The controller instructs the tensor fetcher to load INT4 weights and BFP-formatted activations from external memory into on-chip SRAMs. (2) Once ready, the weight and activation dispatchers stream data to the PE array under the controller's coordination. To maximize data reuse, weights are broadcast across rows, while activations propagate across columns in a systolic pattern. (3) The 8 × 16 PE array then performs BFP-INT or BFP-BFP MACs, depending on input precision and type, following an output-stationary dataflow. Accumulation results are stored in FP32 format to maintain numerical accuracy. (4) After computation, the FP2Half vector unit converts outputs to FP16 format and stores them in the temporary SRAM. (5) For K-type activations in the initial window (detailed in Sec. \ref{sec3-B}), they are forwarded to the online K-offset generator to produce per-channel offsets. For others, this path is bypassed, and this unit is clock-gated to save power. (6) When valid results are available, the BFP converter performs online FP16-to-BFP conversion. (7) Finally, the controller transfers the converted results back to external memory for subsequent operations.

\subsection{Reconfigurable PE Unit}
\begin{figure}[tbp]
\centerline{\includegraphics[width=\linewidth]{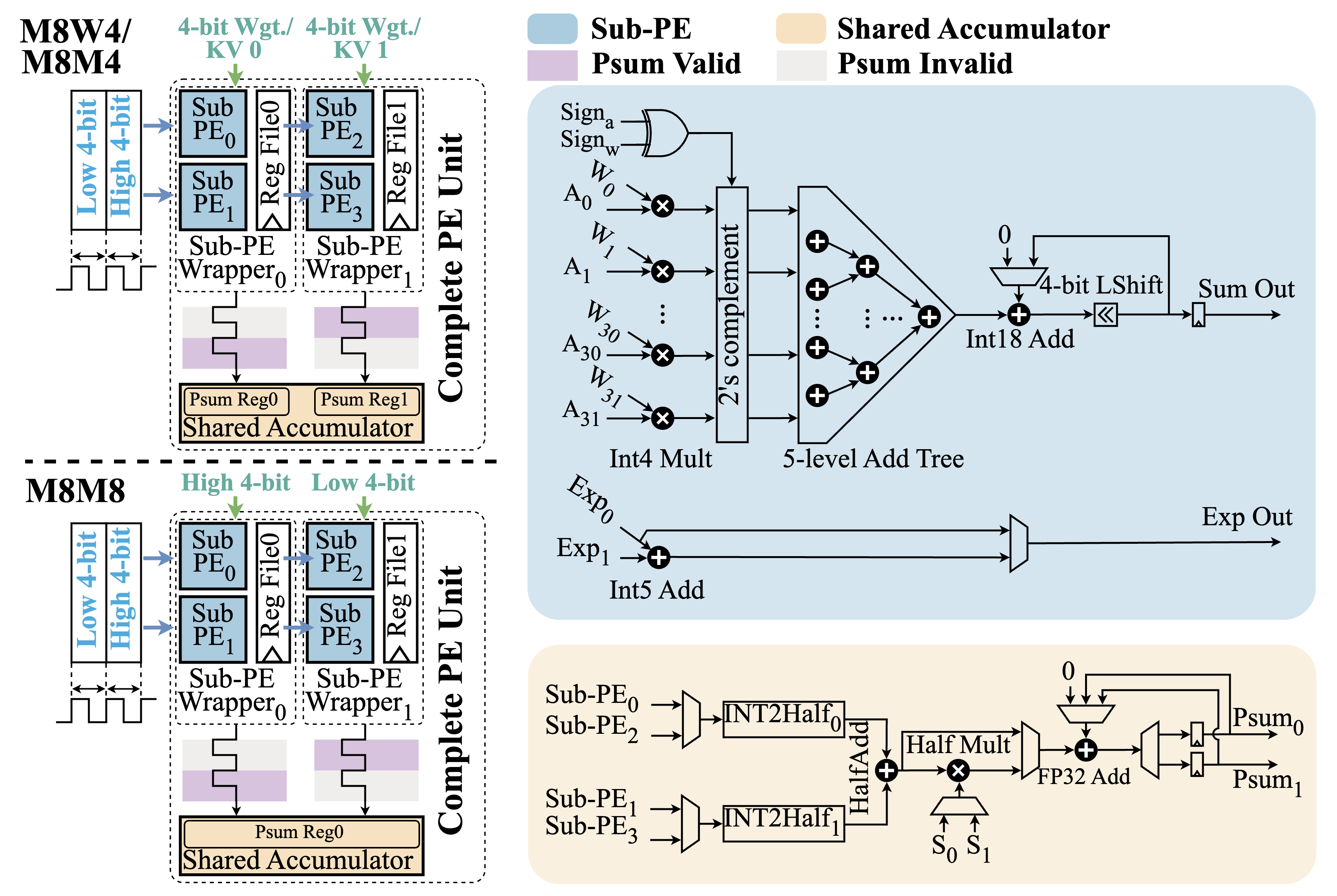}}
\caption{Architecture of the reconfigurable PE unit supporting mixed-format and mixed-precision modes.}
\label{fig13}
\end{figure}

To translate algorithmic optimizations into hardware efficiency, we design a reconfigurable PE unit that flexibly supports mixed-precision and heterogeneous-format computation, as illustrated in Fig. \ref{fig13}. It supports three distinct computation modes: (1) M8W4, which performs MACs between 8-bit-mantissa activations and INT4 weights, (2) M8M4, which performs MACs between activations with 8-bit and 4-bit mantissas, and (3) M8M8, which performs MACs between activations with 8-bit and 8-bit mantissas. The first mode is applied to linear layers, while the latter two are used in attention layers to accommodate mixed-precision activation processing. This unified design maximizes PE utilization and enables fine-grained adaptation to the diverse arithmetic requirements of linear and attention workloads.

The proposed PE unit comprises two sub-PE wrappers, a shared accumulator, and dual register files for buffering activations. Each wrapper integrates two sub-PEs with integer-based compute units capable of executing mixed-precision dot-products on 32 pairs of 4-bit and 8-bit operands in parallel. The shared accumulator is time-multiplexed between wrappers to aggregate cross-group results, amortizing the cost of FP operations while maintaining throughput.

When operating in M8W4 or M8M4 mode, each sub-PE independently performs MACs on 32-element pairs, forwarding partial sums to the shared accumulator for cross-group accumulation. The accumulator then converts integer outputs to FP16 according to their exponents. In M8W4 mode, the FP16 values are further scaled by the group-wise weight factor, while this step is bypassed in M8M4. The systolic propagation of activations allows the accumulator to be shared between two wrappers, improving hardware utilization. In M8M8 mode, both wrappers execute M8M4 computation in parallel, processing the high- and low-4-bit mantissas, respectively. Then, their results are fused in the accumulator to produce the final output.

\subsection{Real-time BFP Converter}
The BFP converter is a key component that transforms various activation types into BFP formats with different mantissa precisions, as described in Sec. \ref{sec3-B}. Tightly integrated with the PE array’s output dataflow, it forms a deeply pipelined path that overlaps computation and conversion, effectively hiding latency. Because different activations use distinct configurations, the converter is carefully designed to handle varying conversion flows efficiently.

\begin{figure}[tbp]
\centerline{\includegraphics[width=\linewidth]{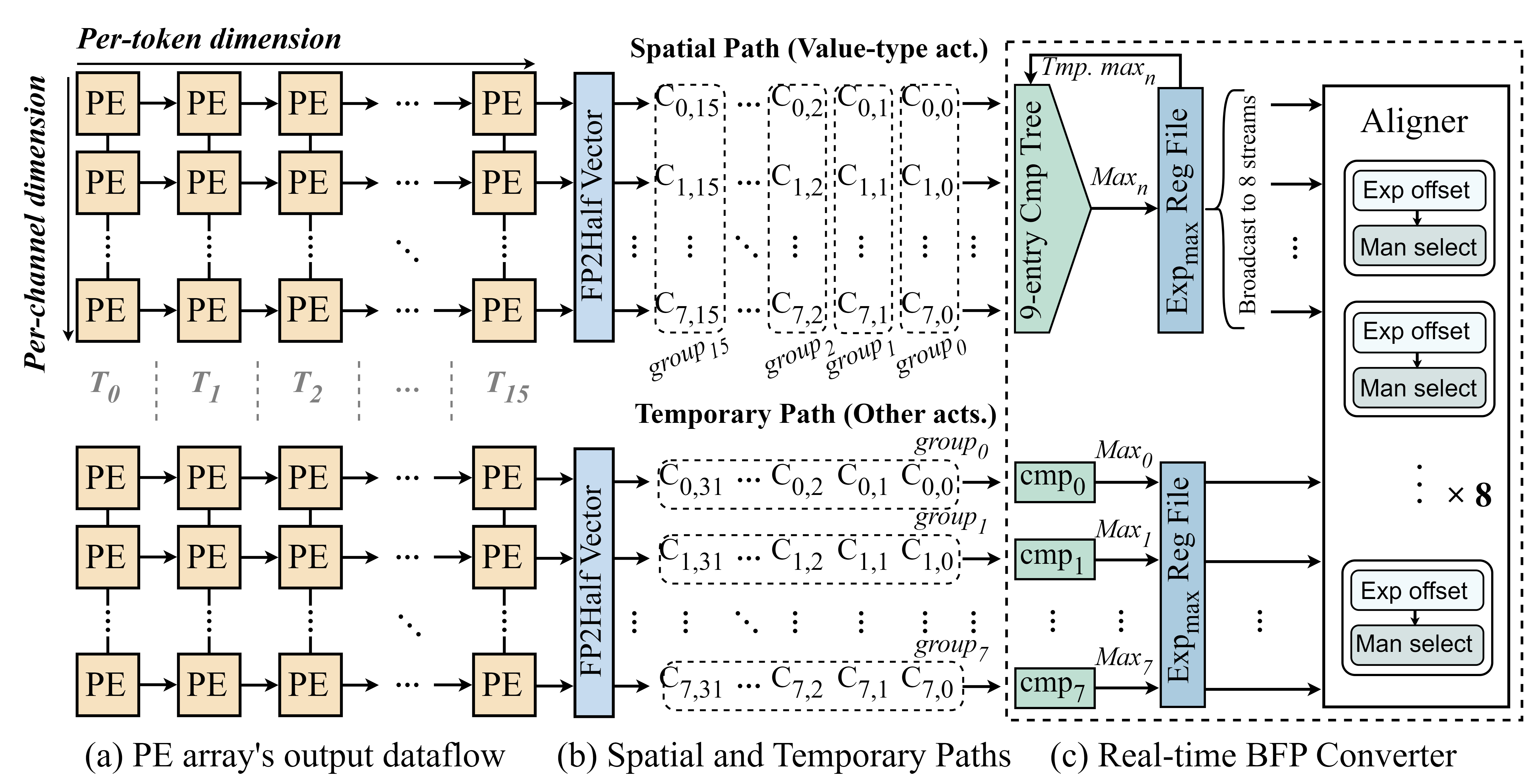}}
\caption{Illustration of (a) output dataflow of the PE array, (b) two data paths supported by the BFP conversion framework, (c) architecture of the BFP converter.}
\label{fig14}
\end{figure}

Fig. \ref{fig14} illustrates two BFP conversion dataflows. All PEs in the same column produce valid results within the same cycle, while outputs from different columns propagate row-wise in a systolic fashion, aligning with the activations’ per-token dimension. For example, results from columns 0 to 15 become valid sequentially from $T_0$ to $T_{15}$. To exploit this pattern, as shown in Fig. \ref{fig14}(b), we design two dedicated conversion paths optimized for different activation types. V-type activations are processed spatially in parallel, while others follow a temporally serialized path.

Specifically, for V-type activations grouped per channel with a group size of 32, as shown in the upper left of Fig.  \ref{fig14}(c), all eight results within a column (e.g., $C_{0,0}$-$C_{7,0}$) are fed into a comparator tree simultaneously to identify the local maximum exponent. Since each group contains 32 elements, four output loops are required to obtain the final maximum exponent. The 32 elements of each group are then divided into four 8-element subgroups, which are forwarded to the aligner along with the derived maximum exponent. Each step processes 8  mantissas in parallel to produce the converted results

In contrast, for per-token grouped activations, each PE row is equipped with a dedicated exponent comparator. As shown in the lower left of Fig. 14(c), results generated in a systolic fashion within a row are streamed directly into the comparator to identify the group’s maximum exponent. After processing 32 results, the final maximum exponent for that row is obtained. Subsequently, the outputs from all eight rows are concurrently forwarded to the aligner, where each row performs mantissa alignment independently in a temporally serialized manner.

Notably, the aligner is shared across the two conversion paths for different activation types, enabling efficient hardware reuse. Furthermore, our proposed real-time BFP converter delivers similar compression effectiveness while achieving a 3.77× reduction in hardware area compared with the online quantizer introduced in the SOTA quantization-based LLM accelerator \cite{hu2025m}.

\subsection{Tiling-aware Dataflow Design}
\begin{figure}[tbp]
\centerline{\includegraphics[width=\linewidth]{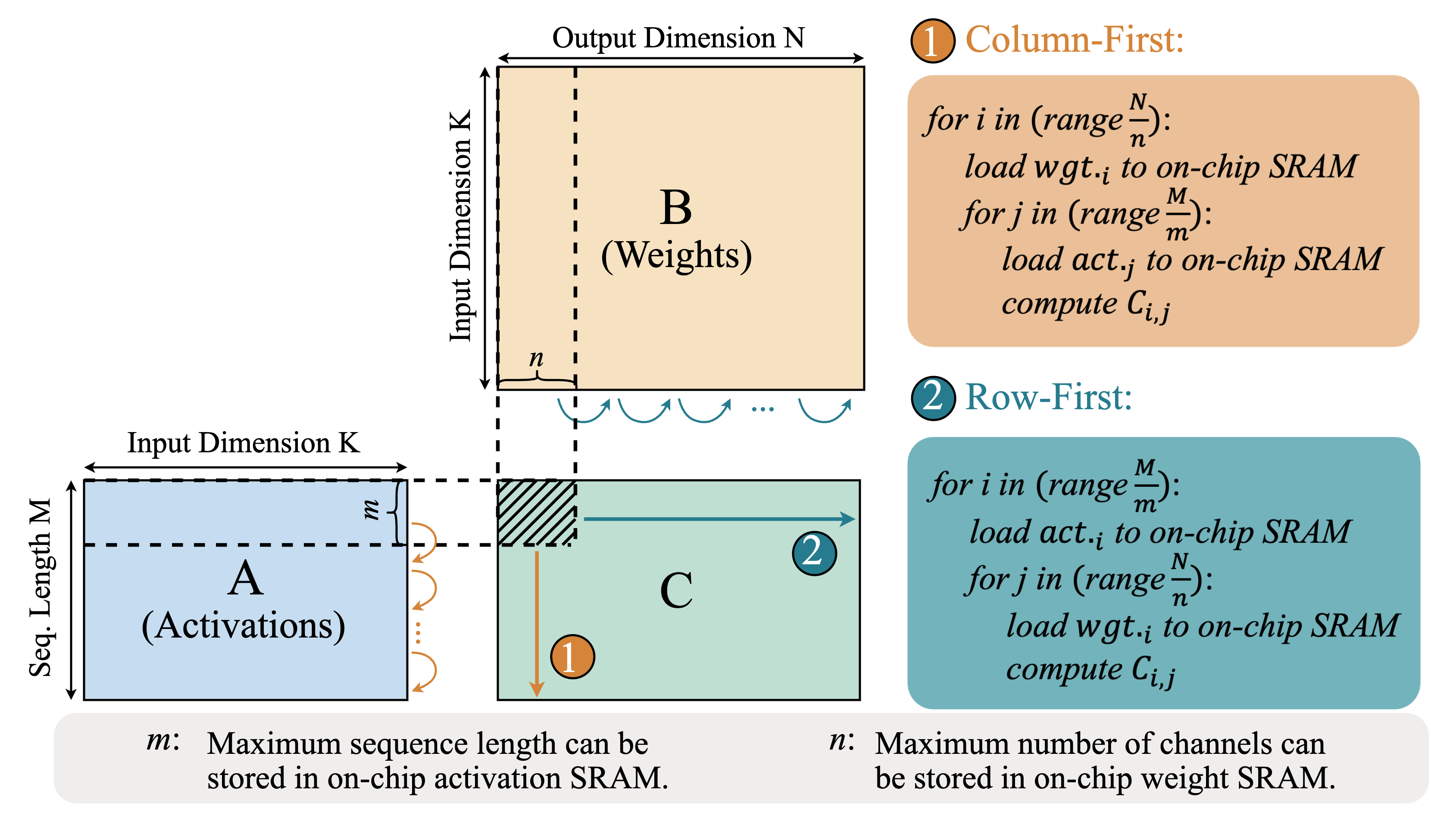}}
\caption{Configurable output dataflows: column-first and row-first modes.}
\label{fig15}
\end{figure}

Due to the limited capacity of on-chip SRAM, tiling strategies are widely adopted to efficiently support large-scale matrix multiplication. As illustrated in Fig. \ref{fig15}, consider multiplying a matrix A of size $M \times K$ with a matrix B of size $K \times N$, resulting in an output matrix C of size $M \times N$. When adopting dataflow 1 (column-first output flow) to compute matrix C, the total number of EMA for matrices A and B is $\frac{N}{n} \times (M \times K) + K \times N$. In contrast, using dataflow 2 (row-first output flow) results in the total number is $\frac{M}{m} \times (K \times N)+(M \times K)$. Taking the linear layer as an example, for a given model, $K$ and $N$ are fixed, while $m$ and $n$ are determined by the on-chip resource allocation and thus remain constant. However, in practical LLM workloads, the number of input tokens can vary substantially, resulting in a wide dynamic range of $M$. Consequently, the choice between the two dataflows directly impacts the EMA cost under different workloads.

Therefore, we design a flexible data generation flow (FDGF) controller that can switch between column-first and row-first dataflows through lightweight configuration. This controller manages the read address order of two data dispatchers from on-chip SRAM and coordinates on/off-chip data exchanges. Specifically, under the column-first dataflow, weights remain in on-chip SRAM until all activations are processed, after which the next weight column is fetched from off-chip memory. Conversely, in the row-first dataflow, activations stay on-chip until computation with all weights is completed before being replaced with new activations. This adaptability enables the architecture to select the most memory-efficient dataflow for a given workload, reducing costly EMA and further improving overall system efficiency.

\section {Evaluation}

\begin{table}[t]
\centering
\caption{Area and power breakdown of the harmonia architecture.}
\label{table0}

\begin{tabular*}{\columnwidth}{
@{\hspace{10pt}}      
l
@{\extracolsep{\fill}}
rr
@{\hspace{10pt}}      
}
\toprule
Component & Area(mm$^2$) & Power(mW) \\
\midrule
PE Array            & 1.4212 & 114.7610 \\
Data Dispatcher     & 0.0729 & 3.8890 \\
On-Chip Memory      & 1.8567 & 180.8730 \\
K-offset Generator  & 0.0129 & 0.4267 \\
BFP Converter       & 0.0009 & 0.0263 \\
Vector Unit         & 0.0027 & 0.1180 \\
Data Collector      & 0.0123 & 0.3240 \\
FDGF Controller     & 0.0016 & 0.1820 \\
\midrule
\textbf{Total}      & \textbf{3.3812} & \textbf{300.6000} \\
\bottomrule
\end{tabular*}
\end{table}

\setlength{\tabcolsep}{1.75pt}
\renewcommand{\arraystretch}{1.0}
\begin{table}[tbp]
\centering
\caption{Perplexity on WikiText2. Lower is better. BFP(x) denotes an x-bit mantissa precision.}
\label{table1}

\resizebox{\linewidth}{!}{
\begin{tabular}{l|ll|ll|llll|l|l}
\toprule
\multirow{2}{*}{Method} 
& \multicolumn{2}{c|}{Linear} 
& \multicolumn{2}{c|}{Atten.} 
& \multicolumn{2}{c|}{Llama} 
& \multicolumn{2}{c|}{Llama-2} 
& \multicolumn{1}{c|}{OPT} 
& \multirow{2}{*}{KV} \\

\cmidrule(lr){2-10}

& Act. & Wgt. 
& Act. & KV 
& 7B & 13B 
& 7B & 13B 
& 6.7B 
& Reduction \\

\midrule
Full      & FP16 & FP16 & FP16 & FP16 & 5.68 & 5.09 & 5.47 & 4.88 & 10.86 & 0.00\% \\
Omniquant & FP16 & INT4 & FP16 & FP16 & 5.77 & 5.17 & 5.59 & 4.95 & 10.96 & 0.00\% \\
Tender    & INT8 & INT8 & FP16 & FP16 & 5.87 & 5.28 & 5.77 & 5.09 & \textbf{10.93} & 0.00\% \\
M-ANT     & INT8 & INT4 & FP16 & FP16 & 5.79 & 5.20 & \textbf{5.57} & 4.96 & 10.98 & 0.00\% \\
FIGNA     & BFP16 & INT4 & FP16 & FP16 & 5.78 & 5.18 & 5.60 & 4.96 & 10.96 & 0.00\% \\
Anda-m8   & BFP8 & INT4 & FP16 & FP16 & 5.79 & 5.19 & 5.60 & 4.97 & 10.96 & 0.00\% \\
Anda-m6   & BFP6 & INT4 & FP16 & FP16 & 5.86 & 5.26 & 5.70 & 5.06 & 11.01 & 0.00\% \\
Anda-m4   & BFP4 & INT4 & FP16 & FP16 & 7.45 & 6.36 & 8.26 & 6.43 & 12.30 & 0.00\% \\
Harmonia  & BFP8 & INT4 & BFP8 & BFP8 & \textbf{5.78} & \textbf{5.18} & 5.59 & \textbf{4.96} & 10.96 & \textbf{42.77\%} \\
Harmonia  & BFP8 & INT4 & BFP8 & BFP4 & 6.19 & 5.51 & 5.90 & 5.20 & 11.69 & \textbf{66.60\%} \\
\bottomrule
\end{tabular}
}
\par\vspace{4pt}
\noindent\begin{minipage}{\linewidth}
\footnotesize\raggedright
\textcolor{blue}{\textit{Note:} Each baseline retains the numerical format and quantization granularity adopted in its original design.}\par
\end{minipage}
\end{table}

\begin{table*}[t]
\centering
\begin{threeparttable}

\caption{Accuracy results on multiple benchmarks from LongBench v1 (context length = 4K). Higher is better.}
\label{table2}

\renewcommand{\arraystretch}{0.55}
\setlength{\aboverulesep}{0.15ex}
\setlength{\belowrulesep}{0.15ex}
\setlength{\tabcolsep}{3pt}

\begin{tabularx}{\textwidth}{
l|l|*{10}{>{\centering\arraybackslash}X}
}
\toprule
\multicolumn{1}{l}{} & &
2wikimqa & repobenchp & hotpotqa &
trec & multinews & multifieldqa &
qasper & qmsum & triviaqa & AVG. \\
\midrule

\multirow{7}{*}{\shortstack{Llama-3.2-\\1B-Instruct}}
& Full & 30.09 & 44.06 & 25.86 & 59.50 & 26.08 & 40.37 & 20.65 & 21.00 & 79.71 & 38.59 \\
& Omniquant & 26.95 & 42.63 & 25.11 & 57.00 & 25.43 & 38.98 & 20.67 & 20.50 & 72.01 & 36.59 \\
& KIVI-q\tnote{*} & \textcolor{blue}{23.50} & \textcolor{blue}{37.07} & \textcolor{blue}{26.19} & \textcolor{blue}{52.00} & \textcolor{blue}{24.47} & \textcolor{blue}{37.08} & \textcolor{blue}{12.41} & \textcolor{blue}{19.38} & \textcolor{blue}{67.70} & \textcolor{blue}{33.31} \\
& \textcolor{blue}{KVQuant-q\tnote{\textdagger}} & \textcolor{blue}{23.45} & \textcolor{blue}{37.47} & \textcolor{blue}{26.14} & \textcolor{blue}{54.00} & \textcolor{blue}{25.02} & \textcolor{blue}{36.03} & \textcolor{blue}{12.81} & \textcolor{blue}{19.23} & \textcolor{blue}{67.55} & \textcolor{blue}{33.52} \\
& MXINT\tnote{+} & 13.03 & 19.73 & 8.31 & 28.50 & 19.65 & 22.52 & 10.55 & 17.41 & 23.14 & 18.09 \\
& \textit{Harmonia-Naïve} & 11.50 & 32.07 & 9.78 & 43.00 & 18.36 & 26.95 & 10.15 & 14.47 & 44.19 & 23.39 \\
& \textbf{Harmonia} & \textbf{26.37} & \textbf{41.05} & \textbf{21.38} & \textbf{56.00} & \textbf{22.70} & \textbf{37.55} & \textbf{20.75} & \textbf{20.57} & \textbf{72.03} & \textbf{35.38} \\
\midrule

\multirow{7}{*}{\shortstack{Llama-3.2-\\3B-Instruct}}
& Full & 34.85 & 48.97 & 38.45 & 68.00 & 26.52 & 46.99 & 35.06 & 20.97 & 87.71 & 45.28 \\
& Omniquant & 31.18 & 52.70 & 34.73 & 64.50 & 25.85 & 46.97 & 32.82 & 21.00 & 86.86 & 44.07 \\
& KIVI-q\tnote{*} & \textcolor{blue}{34.06} & \textcolor{blue}{53.24} & \textcolor{blue}{34.59} & \textcolor{blue}{65.50} & \textcolor{blue}{24.57} & \textcolor{blue}{42.30} & \textcolor{blue}{28.79} & \textcolor{blue}{19.51} & \textcolor{blue}{84.31} & \textcolor{blue}{42.99} \\
& \textcolor{blue}{KVQuant-q\tnote{\textdagger}} & \textcolor{blue}{33.14} & \textcolor{blue}{53.92} & \textcolor{blue}{34.74} & \textcolor{blue}{65.50} & \textcolor{blue}{25.51} & \textcolor{blue}{41.83} & \textcolor{blue}{29.90} & \textcolor{blue}{20.58} & \textcolor{blue}{84.72} & \textcolor{blue}{43.32} \\
& MXINT\tnote{+} & 23.23 & 43.84 & 30.25 & 57.50 & 25.80 & 40.52 & 26.47 & 21.64 & 77.43 & 38.52 \\
& \textit{Harmonia-Naïve} & 29.03 & 42.44 & 26.54 & 60.50 & 25.18 & 39.83 & 26.17 & 20.00 & 81.06 & 38.97 \\
& \textbf{Harmonia} & \textbf{32.76} & \textbf{53.58} & \textbf{36.67} & \textbf{63.00} & \textbf{24.85} & \textbf{44.87} & \textbf{30.49} & \textbf{21.61} & \textbf{89.14} & \textbf{44.11} \\
\midrule

\multirow{7}{*}{\shortstack{Mistral-7B-\\Instruct-v0.3}}
& Full & 31.87 & 54.28 & 36.11 & 73.50 & 26.52 & 48.81 & 32.00 & 21.90 & 88.76 & 45.97 \\
& Omniquant & 26.88 & 53.14 & 33.77 & 74.50 & 27.04 & 46.98 & 30.52 & 22.13 & 89.28 & 44.92 \\
& KIVI-q\tnote{*} & \textcolor{blue}{27.05} & \textcolor{blue}{52.64} & \textcolor{blue}{35.21} & \textcolor{blue}{73.00} & \textcolor{blue}{25.06} & \textcolor{blue}{44.34} & \textcolor{blue}{31.54} & \textcolor{blue}{21.93} & \textcolor{blue}{87.91} & \textcolor{blue}{44.30} \\
& \textcolor{blue}{KVQuant-q\tnote{\textdagger}} & \textcolor{blue}{27.03} & \textcolor{blue}{53.92} & \textcolor{blue}{36.04} & \textcolor{blue}{73.00} & \textcolor{blue}{25.71} & \textcolor{blue}{45.68} & \textcolor{blue}{31.94} & \textcolor{blue}{21.41} & \textcolor{blue}{87.75} & \textcolor{blue}{44.72} \\
& MXINT\tnote{+} & 30.44 & 52.07 & 35.58 & 67.00 & 26.95 & 47.97 & 26.15 & 21.54 & 82.40 & 43.34 \\
& \textit{Harmonia-Naïve} & 30.82 & 48.42 & 38.42 & 70.50 & 27.07 & 42.17 & 28.45 & 21.35 & 87.04 & 43.80 \\
& \textbf{Harmonia} & \textbf{27.34} & \textbf{51.98} & \textbf{34.33} & \textbf{71.50} & \textbf{26.61} & \textbf{45.74} & \textbf{31.65} & \textbf{22.12} & \textbf{88.98} & \textbf{44.47} \\
\midrule

\multirow{7}{*}{\shortstack{Llama-3.1-\\8B-Instruct}}
& Full & 35.23 & 56.29 & 39.04 & 70.00 & 27.21 & 49.67 & 40.24 & 20.96 & 89.01 & 47.52 \\
& Omniquant & 33.70 & 52.15 & 36.19 & 71.50 & 27.19 & 48.95 & 41.70 & 20.98 & 84.75 & 46.35 \\
& KIVI-q\tnote{*} & \textcolor{blue}{31.81} & \textcolor{blue}{36.19} & \textcolor{blue}{36.77} & \textcolor{blue}{69.50} & \textcolor{blue}{25.58} & \textcolor{blue}{44.52} & \textcolor{blue}{38.74} & \textcolor{blue}{19.69} & \textcolor{blue}{83.34} & \textcolor{blue}{42.90} \\
& \textcolor{blue}{KVQuant-q\tnote{\textdagger}} & \textcolor{blue}{32.05} & \textcolor{blue}{36.07} & \textcolor{blue}{37.77} & \textcolor{blue}{70.00} & \textcolor{blue}{25.94} & \textcolor{blue}{44.47} & \textcolor{blue}{38.85} & \textcolor{blue}{20.49} & \textcolor{blue}{83.77} & \textcolor{blue}{43.27} \\
& MXINT\tnote{+} & 28.08 & 35.82 & 34.10 & 59.50 & 27.26 & 39.63 & 29.19 & 21.50 & 83.11 & 39.80 \\
& \textit{Harmonia-Naïve} & 28.29 & 31.95 & 33.04 & 68.50 & 26.63 & 41.69 & 35.49 & 21.48 & 83.96 & 41.23 \\
& \textbf{Harmonia} & \textbf{34.34} & \textbf{55.15} & \textbf{41.45} & \textbf{66.50} & \textbf{25.74} & \textbf{47.61} & \textbf{38.73} & \textbf{20.91} & \textbf{90.26} & \textbf{46.74} \\
\midrule

\multirow{7}{*}{\shortstack{Llama-2-\\13B-chat}}
& Full & 13.21 & 49.90 & 12.49 & 68.50 & 26.56 & 27.20 & 17.09 & 20.89 & 87.75 & 35.95 \\
& Omniquant & 14.70 & 45.95 & 16.31 & 67.00 & 25.81 & 26.94 & 14.57 & 20.09 & 87.31 & 35.41 \\
& KIVI-q\tnote{*} & \textcolor{blue}{14.11} & \textcolor{blue}{47.58} & \textcolor{blue}{13.84} & \textcolor{blue}{64.00} & \textcolor{blue}{22.94} & \textcolor{blue}{27.24} & \textcolor{blue}{16.37} & \textcolor{blue}{20.29} & \textcolor{blue}{86.13} & \textcolor{blue}{34.72} \\
& \textcolor{blue}{KVQuant-q\tnote{\textdagger}} & \textcolor{blue}{13.25} & \textcolor{blue}{48.15} & \textcolor{blue}{13.37} & \textcolor{blue}{64.00} & \textcolor{blue}{23.39} & \textcolor{blue}{28.32} & \textcolor{blue}{17.46} & \textcolor{blue}{20.29} & \textcolor{blue}{86.13} & \textcolor{blue}{34.93} \\
& MXINT\tnote{+} & 15.81 & 39.25 & 12.29 & 47.50 & 22.00 & 26.76 & 12.04 & 16.27 & 56.60 & 27.61 \\
& \textit{Harmonia-Naïve} & 13.06 & 36.26 & 10.30 & 55.50 & 23.77 & 25.38 & 14.38 & 17.65 & 68.48 & 29.42 \\
& \textbf{Harmonia} & \textbf{14.32} & \textbf{44.54} & \textbf{13.79} & \textbf{67.50} & \textbf{25.58} & \textbf{26.30} & \textbf{15.68} & \textbf{20.03} & \textbf{86.84} & \textbf{34.95} \\
\bottomrule
\end{tabularx}

\begin{tablenotes}[flushleft]
\footnotesize
\item[*] \textcolor{blue}{KIVI-q leverages KIVI's KV-cache quantization mechanism, employing the same group size and comparable weight/activation bit-widths as Harmonia.}
\item[\textdagger] \textcolor{blue}{KVQuant-q is a KVQuant-derived aligned variant with comparable weight/activation precisions and aligned group sizes where applicable.}
\item[+] \textcolor{blue}{MXINT uses microscaling-based activation quantization at Harmonia-matched bit widths.}
\end{tablenotes}

\end{threeparttable}
\end{table*}

\subsection{Experimental Setup}
\indent\textbf{Software Implementation:} We implement our proposed algorithm using PyTorch and Hugging Face libraries. The activation quantization (BFP) is implemented with the proposed Harmonia framework and the weight quantization (INT) is realized with the Omniquant \cite{shao2023omniquant} framework. The offline scaling factors are learned on the WikiText2 \cite{merity2016pointer} calibration set by minimizing the BFP-conversion-aware reconstruction error using a straight-through estimator (STE). To demonstrate effectiveness and generality, we benchmark three representative families of open-source  LLMs: OPT \cite{zhang2022opt}, Llama \cite{touvron2023llama}, and Mistral \cite{jiang2023clip},  covering both pretrained models (e.g., OPT-6.7B) and instruction fine-tuned models (e.g., Llama-3.1-8B-Instruct). For pretrained models, we report zero-shot perplexity (PPL) on WikiText2. For instruction fine-tuned models, we evaluate on various tasks from LongBench \cite{bai2024longbench}, including summarization, few-shot learning, document question answering, and code completion. \textcolor{blue}{All experiments are conducted on an RTX 4090 GPU using the official LongBench evaluation settings, which are consistently applied to all evaluated methods.}

\indent\textbf{Algorithm Baselines:} We compare our algorithm against strong baselines that employ either quantization schemes or BFP-based conversion approaches. For PPL evaluation on WikiText2, we consider: (a) Omniquant \cite{shao2023omniquant}, an algorithm that quantizes weights to 4 bits with a group size of 128 while keeping activations in FP16 format. (b) Tender \cite{lee2024tender}, an accelerator that adopts offline activation quantization. (c) M-ANT \cite{hu2025m}, an accelerator utilizing online group-wise activation quantization. (d) FIGNA \cite{jang2024figna}, a hardware framework which enhances numerical accuracy via BFP conversion with extended mantissa precision. (e) Anda \cite{fang2025anda}, an accelerator supporting variable mantissa lengths under BFP-based scheme.

To further evaluate accuracy in more practical scenarios, we evaluate on LongBench. In this setting, we retain baseline (a) and include KIVI \cite{liu2024kivi} and KVQuant \cite{hooper2024kvquant}, two representative algorithms targeting KV-cache quantization to reduce memory footprint. \textcolor{blue}{We construct KIVI-derived and KVQuant-derived aligned variants that retain the core KV-cache quantization mechanisms of the original methods while using Harmonia-comparable payload precisions and aligning the group size at $G=32$ wherever group-wise quantization is applied. These variants are denoted KIVI-q and KVQuant-q, respectively, in Table \ref{table2}.} We also adopt the MXINT \cite{rouhani2023ocp} format for activation quantization as an additional comparison. Moreover, we perform an ablation study by disabling the asymmetric bit allocation strategy and the offline-online hybrid outlier smoothing technique detailed in Sec. \ref{sec3}. This comprehensive setup enables a rigorous assessment of our algorithm.

\indent\textbf{Hardware Implementation:} Harmonia is implemented at the RTL level in SystemVerilog and functionally verified using VCS and Verdi. The prototype is synthesized with Design Compiler using TSMC 28nm process at a clock frequency of 300MHz and a supply voltage of 0.9V. Post-synthesis power analysis is performed in PrimeTime PX using value change dump (VCD) files generated based on the gate-level netlist simulations. On-chip SRAMs are generated using the TSMC-28nm memory compiler, while off-chip memory is modeled as HBM2 with an access energy of 3.9 pJ/bit and a bandwidth of 256 GB/s \cite{jouppi2021ten}.

As summarized in Table \ref{table0}, the synthesis results show that Harmonia occupies 3.38 mm² of silicon area and consumes 300.60 mW of total power. It achieves a peak energy efficiency of 8176 GOPS/W in M8W4/M8M4 mode and 4088 GOPS/W in M8M8 mode, demonstrating consistently high efficiency across different operating configurations.

\indent\textbf{Accelerator Baselines:} To demonstrate the advantages of Harmonia, we compare it against several representative accelerator baselines, each employing distinct strategies for weight-only quantized LLM inference: (a) FP-FP engine, an FP16-based TPU-like\cite{jouppi2017datacenter} accelerator; (b) FP-INT engine, a systolic array with customized FP-INT compute units; (c) FIGNA and FIGNA-C \cite{jang2024figna}, integer-based bit-parallel FP-INT accelerators preserving numerical accuracy; and (d) Anda \cite{fang2025anda}, a bit-serial design supporting variable mantissa precision. All baselines are re-implemented and synthesized under the same technology node, frequency, and voltage as Harmonia for fair comparison. We further develop a cycle-accurate simulator building upon ANT \cite{guo2022ant} and DNNWeaver \cite{sharma2016dnnweaver} to evaluate the energy and performance of all accelerators.

\begin{figure}[tbp]
\centerline{\includegraphics[width=\linewidth]{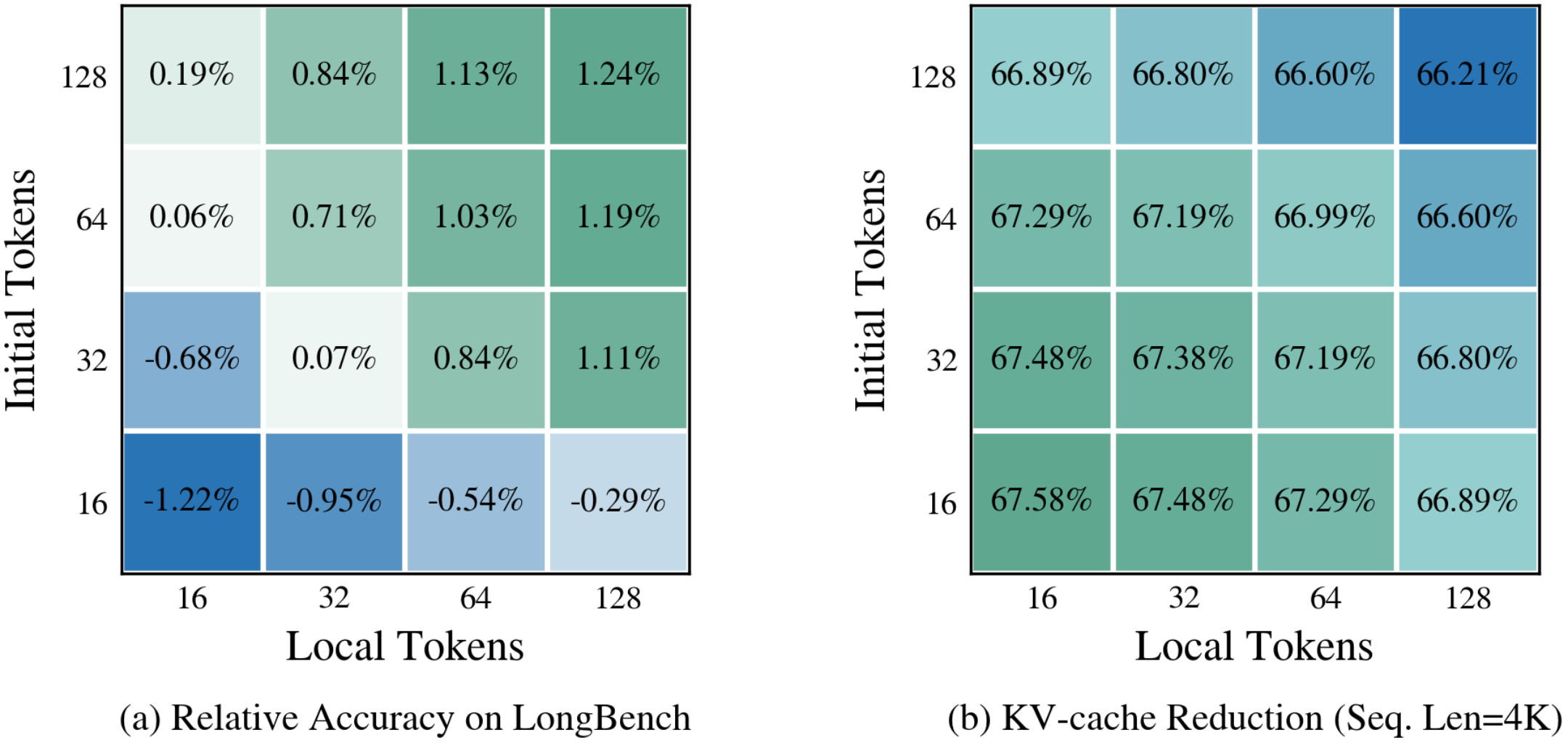}}
\caption{Sensitivity analysis of the initial-local asymmetric bit-allocation strategy on Llama-3.1-8B-Instruct: (a) relative LongBench accuracy with respect to the baseline, and (b) KV-cache reduction relative to the FP16 KV cache.}
\label{fig16}
\end{figure}

\subsection{Model Accuracy Evaluation}
 \indent\textbf{Accuracy Analysis}. For PPL evaluation, we report results under two KV cache precision settings: a conservative 8-bit and an aggressive 4-bit configuration, as shown in Table \ref{table1}. With 8-bit mantissas, our method achieves near-lossless accuracy while reducing KV-cache storage by 42.77\%, whereas all baselines still use FP16 for attention computation and KV-cache storage. Further truncating the mantissas to 4 bits causes a moderate accuracy drop but yields a 66.60\% reduction in storage, representing a practical trade-off between accuracy and memory efficiency.

We further evaluate five instruction-tuned models ranging from 1B to 13B parameters. To better reflect real-world scenarios, their performance is assessed on the LongBench benchmark suite, as summarized in Table \ref{table2}. \textcolor{blue}{Our approach incurs a 0.94\% average accuracy drop across five evaluated models compared with the weight-only quantization baseline}, while delivering significant gains in both computation and memory efficiency. Notably, for Llama-3.2-3B-Instruct and Llama-3.1-8B-Instruct, it even surpasses the baseline accuracy, demonstrating its advantage. \textcolor{blue}{We also observe some model-dependent variation, with relatively larger degradation on smaller models, suggesting that they may be more sensitive to BFP compression. For such models, the BFP-m8 initial- and local-token windows can be enlarged to trade part of the KV-cache compression ratio for improved accuracy. Moreover, compared with KIVI-q and KVQuant-q, which are KIVI-derived and KVQuant-derived aligned variants, respectively, Harmonia achieves average accuracy improvements of 3.76\% and 2.98\%, respectively. Harmonia also consistently outperforms MXINT across all five models, indicating better accuracy at equivalent bit widths. The ablation study further shows a 5.77-point average gain over the naive version}, confirming the effectiveness of our algorithm in mitigating accuracy degradation under aggressive compression.

\textcolor{blue}{\indent\textbf{Sensitivity Analysis}. To evaluate the sensitivity of the asymmetric bit allocation strategy, we vary the numbers of initial and local tokens over $\{16, 32, 64, 128\}$ on Llama-3.1-8B-Instruct. As shown in Fig. \ref{fig16}, enlarging these high-precision windows generally improves the relative LongBench accuracy but reduces the achievable KV-cache reduction, revealing a clear trade-off between accuracy and storage efficiency. For models more sensitive to KV-cache compression, these windows can be moderately enlarged to prioritize accuracy. }

\begin{figure*}[t]
\centering
\includegraphics[width=\textwidth]{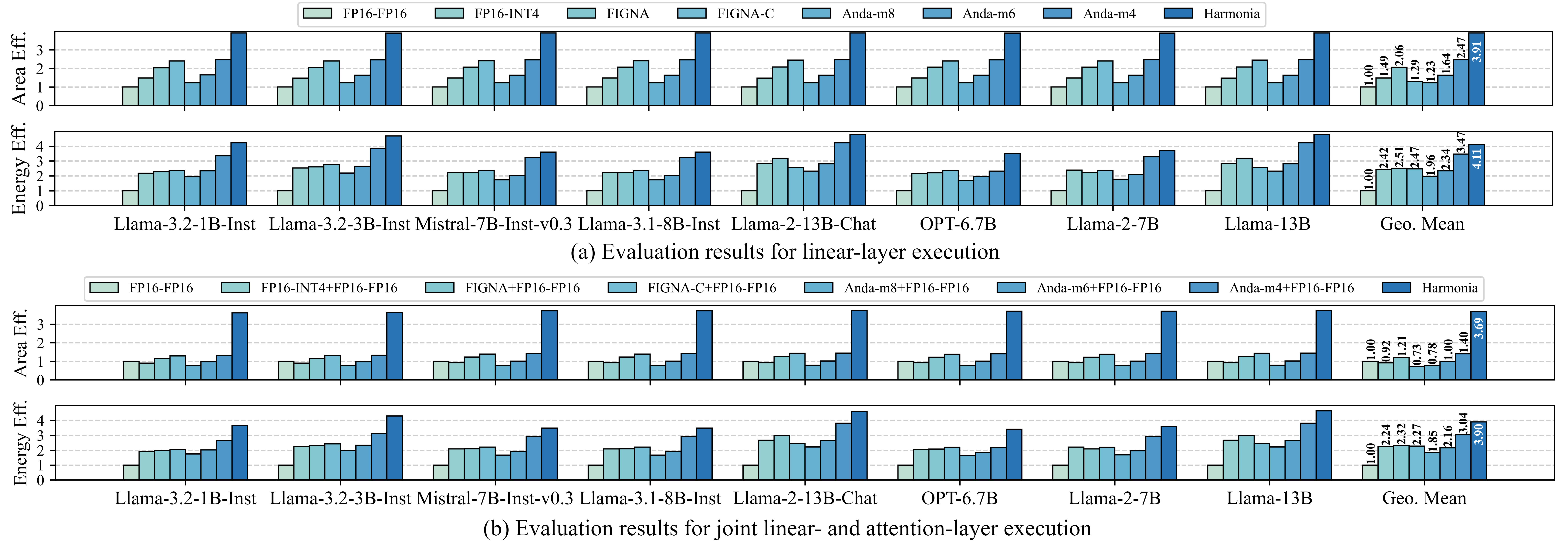}
\caption{Area efficiency and energy efficiency comparison across accelerators in (a) linear-layer execution and (b) joint linear- and attention-layer execution.}
\label{fig17}
\end{figure*}

\begin{figure}[tbp]
\centerline{\includegraphics[width=\linewidth]{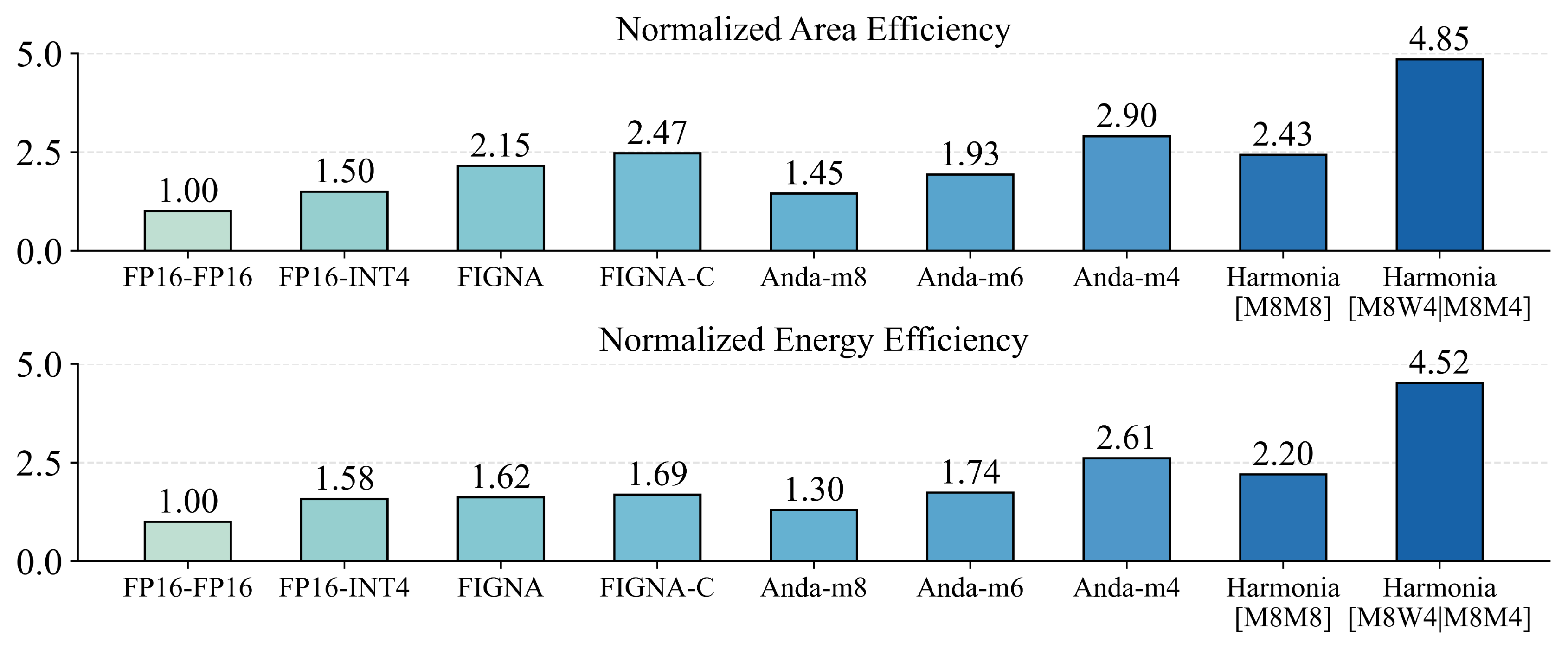}}
\caption{PE-level comparison of area efficiency and energy efficiency, with all results normalized to the FP16-FP16 baseline.}
\label{fig18}
\end{figure}

\subsection{PE-level Hardware Evaluation}
We quantitatively compare the PE unit of Harmonia with hardware baselines in terms of area efficiency and energy efficiency. To ensure fairness, identical dot-product workloads extracted from real-world LLM inference are executed on each PE. The results, shown in Fig. \ref{fig18}, report area efficiency (TOPS/mm²) and energy efficiency (TOPS/W), respectively. Here, Anda-m(x) denotes the Anda PE configured with an x-bit mantissa precision.

Harmonia supports three operating modes: M8W4, M8M4, and M8M8, which differ in their internal parallelism and resource utilization. In the M8W4/M8M4 mode, two sub-PE wrappers in the PE independently perform 64 MACs in parallel, yielding a total throughput of 128 MACs per PE. In contrast, in the M8M8 mode, the two wrappers collaboratively execute 64 MACs over the same number of clock cycles. Consequently, the energy efficiency and area efficiency in the M8M8 mode are half of those in the M8W4/M8M4 mode.

\begin{figure*}[t]
\centering
\includegraphics[width=\textwidth]{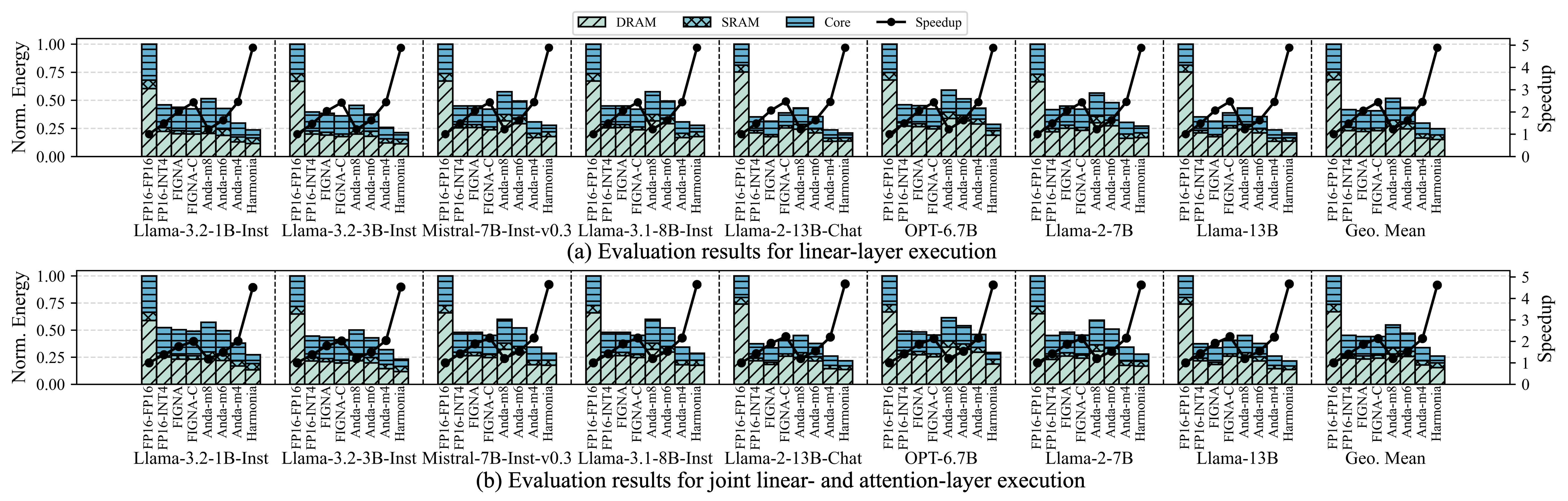}
\caption{Speedup and energy breakdown comparison across accelerators in (a) linear-layer execution and (b) joint linear- and attention-layer execution.}
\label{fig19}
\end{figure*}

\begin{figure}[htbp]
\centerline{\includegraphics[width=\linewidth]{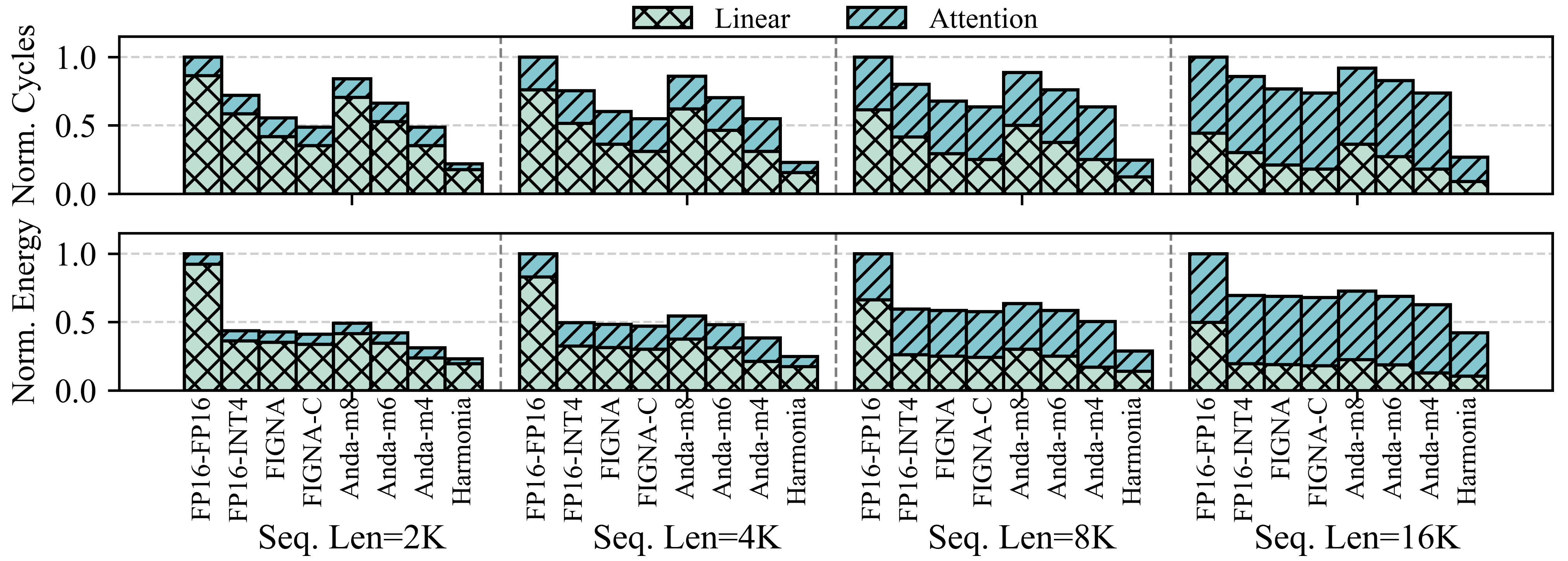}}
\caption{Speedup and energy comparison across accelerators under varying sequence lengths.}
\label{fig20}
\end{figure}

As shown in Fig. \ref{fig18}, in M8W4/M8M4 mode, Harmonia achieves 1.67-4.85$\times$ higher area efficiency and 1.73-4.52$\times$ higher energy efficiency than baseline designs. In M8M8 mode, its area efficiency is slightly below FIGNA-C and Anda-m4, while energy efficiency remains comparable. These results demonstrate the scalability and flexibility of Harmonia across different precision configurations, effectively balancing among throughput, area and energy. This reconfigurable PE unit thus forms the foundation for system-level improvements in performance and efficiency.

\subsection{System-level Hardware Evaluation}
For a fair iso-area comparison, each baseline accelerator is configured with a PE array occupying the same silicon area as that of Harmonia, while maintaining identical on-chip SRAM capacity and operating frequency. \textcolor{blue}{We separately evaluate the prefill and decode stages to account for their distinct computation patterns and memory-access characteristics.}

\indent\textbf{Prefill Stage}. For the prefill-stage evaluation, we deploy various LLM workloads with a batch size of 1 and a sequence length of 2048 tokens. We evaluate area efficiency, energy efficiency, speedup, and energy breakdown in two scenarios: (1) GEMM operations in linear layers only, and (2) GEMM executed jointly in both linear and attention layers. Since all accelerators except the FP16-FP16 engine and Harmonia lack native support for attention-layer computation, an auxiliary FP16-FP16 unit is integrated to handle attention layers for functional completeness. The linear-layer evaluation results are presented in Fig. \ref{fig17}(a) and Fig. \ref{fig19}(a), while Fig. \ref{fig17}(b) and Fig. \ref{fig19}(b) show results for joint linear-attention execution.

Results from the linear-layer evaluation show that Harmonia achieves a 1.58-3.91$\times$ higher area efficiency and 1.18-4.11$\times$ improvement in energy efficiency on average, while delivering up to 4.89$\times$ speedup and reducing total energy consumption by 4.15$\times$ compared with the FP16-FP16 baseline. When GEMM operations in both linear and attention layers are considered, Harmonia achieves up to 5.05$\times$ and 3.90$\times$ improvements in area efficiency and energy efficiency, respectively. Meanwhile, it delivers an average 2.17-4.62$\times$ speedup and reduces average energy consumption by 1.27-3.99$\times$ compared with baselines.

To further validate the versatility of Harmonia, we evaluate its performance across varying sequence lengths. Fig. \ref{fig20} presents the normalized execution cycles and energy of Harmonia and baseline accelerators using the Llama-3.2-3B-Instruct model with sequence lengths from 2K to 16K. As the sequence length grows, the attention layer gradually becomes dominant. With unified optimization for both linear and attention operations, Harmonia consistently outperforms all baselines, achieving significant reductions in execution cycles and energy consumption. Specifically, from 2K to 16K tokens, Harmonia delivers 2.50-4.14× speedup and 1.54-3.35× energy reduction, demonstrating strong scalability and efficiency across varying workloads.

\textcolor{blue}{\indent\textbf{Decode Stage}. We further evaluate decode-stage performance and energy consumption using the Llama-3.1-8B-Instruct model with a batch size of 1. For the linear layers, we perform single-token GEMV using all evaluated accelerator designs and report the normalized execution cycles and energy breakdown. For the attention layers, we process one query token over context lengths ranging from 2K to 16K. The attention-layer evaluation includes the FP16-FP16 baseline and two Harmonia configurations, namely Harmonia-M8 and Harmonia-Mix. Harmonia-M8 stores the entire KV cache in BFP-m8, whereas Harmonia-Mix retains the first 32 and most recent 64 KV-cache tokens in BFP-m8 and stores the remaining tokens in BFP-m4.}

\begin{figure}[tbp]
\centerline{\includegraphics[width=\linewidth]{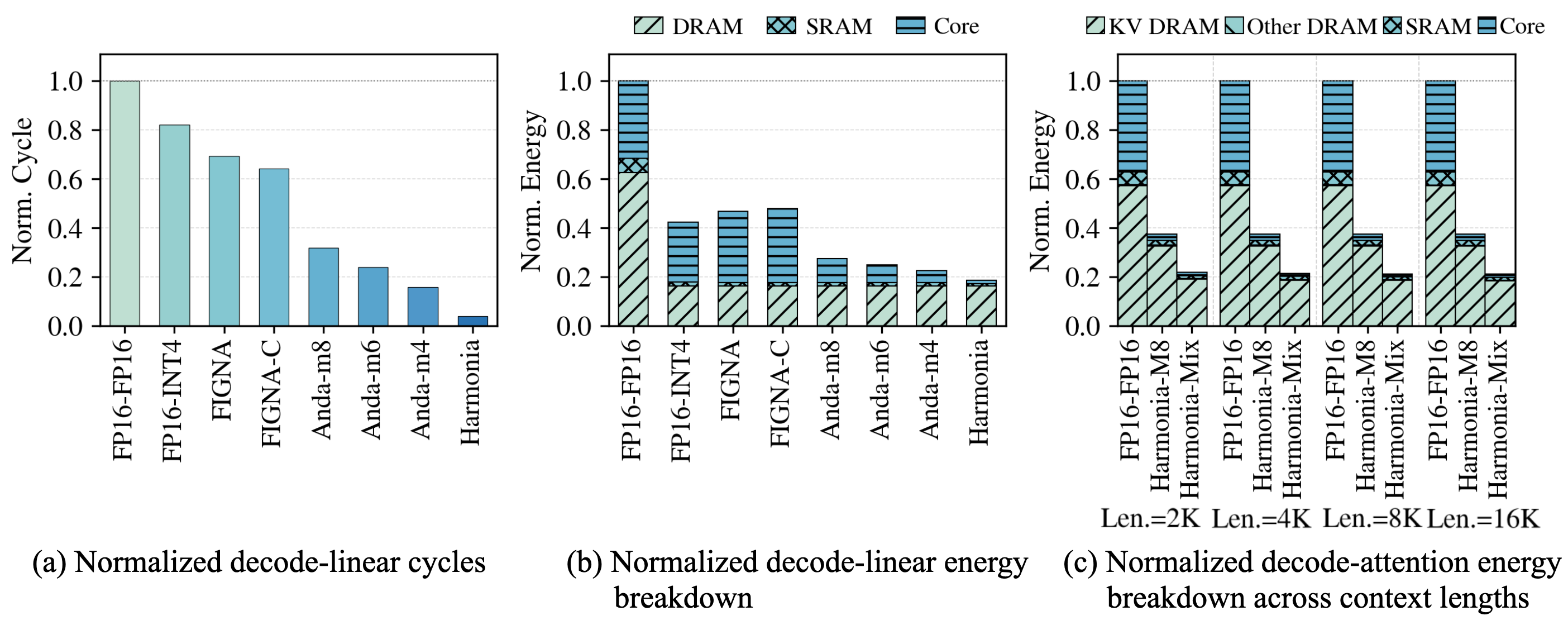}}
\caption{Decode-stage hardware evaluation results on Llama-3.1-8B-Instruct. (a) Normalized execution cycles and (b) normalized energy breakdown for linear layers. (c) Normalized energy breakdown for attention layers across context lengths from 2K to 16K.}
\label{fig21}
\end{figure}

\textcolor{blue}{As shown in Fig.~\ref{fig21}(a) and (b), Harmonia achieves a 4.00--25.02$\times$ speedup and reduces total energy consumption by 1.21--5.35$\times$ over the evaluated baselines. During batch-size-1 decoding, the single token occupies only one PE-array row. Harmonia therefore achieves a theoretical row-level utilization of 12.5\%, compared with 1.54\%--6.25\% for the baselines, which retain their iso-area square-array configurations with larger row counts. The speedup thus reflects both Harmonia's efficient mixed-format BFP computation and its higher row occupancy. Harmonia does not employ a specialized array mapping to address GEMV underutilization, but instead targets computation and memory overheads through BFP-based activation representation, reconfigurable mixed-format computation, and KV-cache compression. When multiple decode requests are processed concurrently, their newly generated tokens can be processed together in the linear layers, increasing PE-array row occupancy. For Harmonia, batch sizes of 4, 8, and 16 yield theoretical row-level utilizations of 50\%, 100\%, and 100\%, respectively, showing that small-batch request processing can rapidly recover PE-array parallelism.}

\textcolor{blue}{As shown in Fig. \ref{fig21}(c), Harmonia-M8 and Harmonia-Mix reduce the total decode-stage attention energy by 2.66$\times$ and 4.56--4.71$\times$, respectively. Compared with the full-precision KV cache, they reduce KV-cache DRAM energy by 42.77\% and 66.60\%--67.62\%, respectively. These savings arise from BFP-based KV-cache compression and the lower core and on-chip SRAM energy enabled by reduced-precision BFP computation. As the context length increases, a larger proportion of the KV cache in Harmonia-Mix is stored in BFP-m4, leading to progressively greater energy savings.}

\textcolor{blue}{Taken together, these results demonstrate Harmonia's effectiveness in both the prefill and decode stages.} The performance gains mainly stem from two key factors: (1) \noindent\textbf{Reconfigurable compute core design:} Harmonia features a reconfigurable PE array that unifies mixed-type and mixed-precision computation in linear and attention layers. It flexibly composes mixed-precision operations from 4-bit basic building blocks and time-multiplexes shared accumulators between wrappers, achieving high utilization with minimal area cost. Compared with FIGNA’s wide-bit parallel MAC units and Anda’s bit-serial design, Harmonia strikes a balance among area, power, and throughput. (2) \noindent\textbf{Compressed activation representation:} Harmonia adopts a mixed-precision, mantissa-compressed BFP format for activation conversion and storage. Using an asymmetric bit-allocation strategy that truncates mantissas to 8 or 4 bits, it greatly reduces memory access overhead and further improves overall energy efficiency.

\section {Related Work and Discussion}
\indent\textbf{LLM accelerators.} To mitigate the high computational and memory costs of LLMs, many accelerators have been proposed, mainly focusing on weight-only or weight-activation quantization. For weight-only quantization, \cite{jang2024figna}, \cite{fang2025anda} convert FP activations to the BFP format to enable integer-based FP-INT computation, while \cite{park2025figlut}, \cite{mo2025lut} employ LUT-based arithmetic. However, these designs primarily optimize linear layers and ignore the costly FP-FP computations in attention modules. Later efforts \cite{chen2025bitmod}, \cite{guo2023olive}, \cite{lee2024tender}, \cite{hu2025m} quantize activations to further reduce overhead, but activation quantization is challenging: offline quantization often degrades accuracy, while dynamic quantization requires expensive runtime scaling and offset computation. In contrast, Harmonia extends the BFP format to activations in both linear and attention layers, which together dominate over 90\% of LLM computation, reducing the cost of activation compression compared with dynamic quantization and striking a balance between model accuracy and hardware efficiency.

\indent\textbf{KV cache optimization.} \textcolor{blue}{During autoregressive generation, frequent access to the KV cache causes heavy memory and bandwidth pressure, especially in long-context scenarios. Existing methods \cite{yue2024wkvquant}, \cite{su2025accurate} reduce this overhead through KV-cache-specific quantization but introduce runtime overhead. Oaken~\cite{kim2025oaken} further mitigates this runtime overhead with an online-offline lightweight KV-cache integer quantization strategy and an efficient memory management unit. It profiles outlier thresholds offline and uses them online to guide outlier-aware group-shift quantization, reducing the cost of handling dynamic KV-cache distributions during inference. TurboQuant~\cite{zandieh2025turboquant} achieves more aggressive KV-only compression through online vector quantization. Harmonia instead trades some KV-only compression aggressiveness for lower online codec complexity, allowing its mixed BFP-m4/BFP-m8 KV cache to be directly consumed by the BFP attention datapath and integrated with low-bit computation in both linear and attention layers. Moreover, this approach is orthogonal to other techniques, such as eviction \cite{zhang2023h2o}, \cite{liu2023scissorhands}, and sliding window \cite{duanmu2024skvq}, \cite{xiao2023efficient}, allowing further compression when combined.}

\section {Conclusion}
This work presents Harmonia, an algorithm-hardware co-design framework that converts activations in both linear and attention layers into the BFP format, and integrates efficient reconfigurable hardware units to support mixed-precision BFP-INT and BFP-BFP operations for weight-only quantized LLMs. We achieve this by carefully selecting a BFP configuration with a group size of 32 and a mantissa length of 8 to preserve model accuracy. Building on this foundation, we introduce an asymmetric bit-allocation and offline-online hybrid outlier smoothing technique, compressing the KV cache to about 33.0\% of its original size, reducing both memory footprint and access overhead. To fully leverage algorithmic benefits, we design complementary hardware components, including a reconfigurable mixed-precision PE array, a real-time BFP converter, and a flexible tiling-aware dataflow. Comprehensive evaluations demonstrate that Harmonia achieves a 3.08$\times$ speedup, a 3.84$\times$ improvement in area efficiency, and a 2.03$\times$ enhancement in energy efficiency on average over prior works. These results highlight Harmonia's adaptability for efficiently handling diverse LLM workloads.

\section {Acknowledgment}
\textcolor{blue}{We declare that the generative AI tool, ChatGPT (GPT-5), was used solely for language polishing and grammar correction in the preparation of this manuscript. It did not contribute to the research content or methodology. All AI-assisted revisions were carefully reviewed and verified by the authors. We also thank the reviewers for their valuable comments and constructive suggestions, which helped improve the quality of this work.}

\bibliographystyle{IEEEtran}
\bibliography{refs}

\makeatletter
\def\@IEEEBIOskipN{1\baselineskip}
\makeatother



\end{document}